# Pseudonym Scheme Based on Hybrid Certificates for Security Credential Management System in Vehicular Communications

Abel C. H. Chen, *Senior Member, IEEE*, F. J. Hwang, Yu-Chih Wei, Chin-Chen Chang, *Fellow, IEEE*, and Bon-Yeh Lin

***Abstract*—In recent years, the Institute of Electrical and Electronics Engineers (IEEE) and the European Telecommunications Standards Institute (ETSI) have developed a series of security communication standards for vehicular communications. These standards include mechanisms such as the Security Credential Management System (SCMS) and Butterfly Key Expansion (BKE) to protect vehicle privacy. However, these standards are mainly based on the Elliptic-Curve Cryptography (ECC), which may be vulnerable to attacks from quantum computing in the future. In response to this potential risk, this study proposes a hybrid certificate that combines the ECC with Post-Quantum Cryptography (PQC). This approach enables infrastructure systems to be built on cryptographic foundations that are more resilient to quantum-based attacks. Furthermore, this study presents a generalized pseudonym scheme that is compatible with various cryptographic algorithms for generating pseudonym certificates. This design aims to eliminate the possibility of inferring any correlation between the public key in a pseudonym certificate and that in an enrollment certificate. This study also conducts a comprehensive performance evaluation of the RSA, ECC, and PQC algorithms, particularly those standardized by the National Institute of Standards and Technology (NIST). The comparison considers factors such as message length and computation time. Based on the findings, this study recommends suitable pseudonym schemes that adopt hybrid certificates for secure and efficient use in vehicular communications.**

## I. Introduction

WITH the widespread adoption of information and communication technologies and artificial intelligence applications, the security of Vehicle-to-Everything (V2X) communications has emerged as a critical topic of global concern. In recent years, the Institute of Electrical and Electronics Engineers (IEEE) has proposed the Security Credential Management System (SCMS) [1],[2], while the European Telecommunications Standards Institute (ETSI) has introduced the Cooperative Intelligent Transport Systems (C-ITS) Security Credential Management System (CCMS) [3],[4]. These efforts have led to the development of relevant standards for vehicular communications, such as certificate structure specifications defined in IEEE 1609.2 [5] and ETSI TS 103 097 [6], and certificate request procedures outlined in IEEE 1609.2.1 [7] and ETSI TS 102 941 [8]. To ensure interoperability, working groups from both IEEE and ETSI have referenced each other's designs and architectures, with the expectation that many structures and processes will eventually align. For instance, Version 2.1.1 of ETSI TS 103 097 defines its certificate structure primarily based on IEEE 1609.2 [5],[6], while Version 2.2.1 of ETSI TS 102 941 incorporates the Butterfly Key Expansion (BKE) mechanism as defined in IEEE 1609.2.1 [7],[8]. To maintain terminological clarity, this study adopts the terminology and definitions used by IEEE. Specifically, terms such as "Authorization Certificate" and "Anonymous Certificate" are consistently referred to as "Pseudonym Certificate (PC)." Furthermore, this study builds upon the foundations of IEEE 1609.2 and IEEE 1609.2.1 to propose improvements to existing certificate structures and processes.

Since vehicular communications may involve the tracking of individual vehicle movements and behaviors, they present significant privacy concerns [9],[10],[11]. To address these concerns, IEEE 1609.2.1 introduces the BKE process based on the properties of Elliptic-Curve Cryptography (ECC). This process involves expanding a caterpillar public key into a cocoon public key, and subsequently into a butterfly public key, which is then embedded in the pseudonym certificate. This design prevents other entities in the vehicular communications, including the Certificate Authority (CA) and Registration Authority (RA), from inferring the certificate holder's identity through the butterfly public key contained in the pseudonym certificate [7]. However, the current BKE mechanism is tightly coupled with ECC and may not be directly applicable to other cryptographic approaches. Its design lacks the flexibility to migrate to alternative cryptographic primitives. Furthermore, ECC does not offer

Abel C. H. Chen and Bon-Yeh Lin are with the Information & Communications Security Laboratory, Chunghwa Telecom Laboratories, Taoyuan 32661, Taiwan (e-mail: chchen.scholar@gmail.com; bylin@cht.com.tw). *(Corresponding author: Abel C. H. Chen)*

F. J. Hwang is with Department of Business Management, National Sun Yat-sen University, Kaohsiung 804201, Taiwan (e-mail: fengjanghwang@gmail.com).

Yu-Chih Wei is with the Department of Information and Finance Management, National Taipei University of Technology, Taipei 106344, Taiwan (e-mail: vickrey@mail.ntut.edu.tw). *(Co-Corresponding author: Yu-Chih Wei)*

Chin-Chen Chang is with the Department of Information Engineering, Feng Chia University, Taichung 407802, Taiwan (e-mail: alan3c@gmail.com).

resistance to attacks from quantum computing and is potentially vulnerable to future quantum-based cryptanalysis.

With the continued advancement of quantum computing technologies, there is growing concern that widely used cryptographic algorithms such as RSA and ECC may become vulnerable to quantum-based attacks [12]. In response to this emerging threat, the National Institute of Standards and Technology (NIST) finalized several Post-Quantum Cryptography (PQC) standards in 2024. These include the Module-Lattice-Based Key Encapsulation Mechanism (ML-KEM) [13], the Module-Lattice-Based Digital Signature Algorithm (ML-DSA) [14], and the Stateless Hash-Based Digital Signature Algorithm (SLH-DSA) [15]. In addition to standardization efforts, both the United States and the European Union have published proposed roadmaps and timelines for migrating to PQC [16],[17]. As a result, evaluating current cryptographic methods and transitioning to quantum-safe alternatives is regarded as an essential step for ensuring long-term security. However, in the context of vehicular communication applications, there are not only stringent security and privacy requirements but also strict constraints on packet length. If the length of a message exceeds the allowable packet length, transmission may be infeasible. These limitations highlight the importance of carefully selecting cryptographic schemes that balance quantum resistance with practical implementation constraints in resource-constrained environments such as vehicular communications.

This study identifies the following key challenges in the design of SCMS for vehicular communications:

- **Packet Length Limitation**: According to the IEEE 1609.3 standard, the default maximum length for a Wireless Access in the Vehicular Environment (WAVE) Short Message (WSM) is 1400 bytes [18]. Therefore, any packet transmitted by a vehicle must not exceed this length limit. Packets that exceed this threshold cannot be transmitted, imposing a strict constraint on the length of cryptographic content included in each message.
- **Signature Generation and Verification Efficiency**: In North American vehicular communication Proof-of-Concept (POC) deployments, On-Board Units (OBUs) are required to transmit a Secure Protocol Data Unit (SPDU) containing a Basic Safety Message (BSM) every 100 milliseconds [19]. Each SPDU must include a signature. Additionally, OBUs must be capable of receiving SPDUs from surrounding OBUs or Road-Side Units (RSUs) and verifying the included signatures in real time. Therefore, achieving high-performance signature generation and verification is essential to meet the real-time communication demands of vehicular environments.
- **Security Level (SL)**: The NIST defines security levels to evaluate the strength of PQC algorithms [20]. However, higher security levels typically result in larger public keys and certificate lengths, as well as longer signature generation and verification times. As such, it is critical to design solutions that strike a balance between cryptographic security, packet length constraints, and processing efficiency.
- **Vehicle Privacy**: Since most vehicles in vehicular communications are privately owned, preserving individual privacy is a significant concern [7]. In this study, vehicle privacy is defined as the inability to infer a link between the public key in a PC and the public key in an Enrollment Certificate (EC). To protect personal and vehicle-related information, one of the key challenges is to design pseudonym certificates that are compatible with PQC schemes while ensuring unlinkability and privacy.

In response to the aforementioned challenges, this study proposes a pseudonym scheme based on hybrid certificates for applications within SCMS. The scheme is first categorized according to vehicle communication types: Vehicle-to-Vehicle (V2V), Vehicle-to-Infrastructure (V2I), and Infrastructure-to-Infrastructure (I2I). Given the 1400-byte packet length limitation defined in the IEEE 1609.3 standard, the V2V component adopts a hybrid certificate structure that combines ECC and PQC. In contrast, since V2I and I2I communications are not subject to the same packet length constraint, pure PQC-based certificates are employed to secure infrastructure communications. To further enhance vehicle privacy, this study proposes a generalized pseudonym scheme that is not limited to ECC and can support pseudonym certificate generation using any combination of cryptographic algorithms. This design ensures unlinkability between the public keys in PCs and those in ECs, thereby preserving user privacy.

The main contributions of this study are summarized as follows:

- **Proposing ECC and PQC Hybrid Certificates**: A certificate and message structure tailored to meet the 1400-byte packet length constraint is designed for V2V communications. The hybrid certificate contains a PQC-based signature issued by the CA, along with an ECC-based public key associated with the end entity (EE), such as an OBU or a RSU. The proposed scheme is primarily designed to ensure quantum security by adopting pure PQC certificates with long validity requirements within the infrastructure environment. These include Root Certificate Authority (RCA), Intermediate Certificate Authorities (ICAs), Enrollment Certificate Authority (ECA), Pseudonym Certificate Authority (PCA), and Registration Authority (RA) certificates, as well as ECs for EEs. In contrast, PCs based on hybrid certificates used by EEs have inherently short lifetimes. Even in the event of future advancements in quantum computing capabilities, their security can be maintained by further reducing their validity periods as needed.
- **Proposing a Generalized Pseudonym Scheme**: A cryptographic framework is proposed that allows pseudonym certificates to be generated using any type of cryptographic method. This approach ensures that the public key in a PC cannot be linked to the corresponding public key in an EC.
- **Comparing the Performance of RSA, ECC, and PQC Algorithms in the Proposed Methods**: This

study experimentally evaluates the performance of RSA, ECC, and PQC algorithms in the context of hybrid certificates. Metrics such as message length and computation time are analyzed to identify the most suitable cryptographic solution for vehicular communications, thereby providing a reference for future deployment. Furthermore, this study assesses the computational requirements of the proposed scheme under different Levels of Service (LOS). The results show that the proposed scheme is able to satisfy the required computational demands and outperforms the methods specified in the IEEE 1609.2 and IEEE 1609.2.1 standards.

The structure of this paper consists of five sections. Section II reviews related literature, including SCMS for vehicular communications and PQC methods. Section III presents the proposed pseudonym scheme based on hybrid certificates, detailing the methodology and underlying principles. Section IV discusses the experimental results and provides analysis, comparing performance in terms of message length and computation time. Finally, Section V concludes the study by summarizing the key contributions and outlining potential directions for future research.

## II. Literature Reviews

This section begins by presenting the architecture and data structures of the SCMS as defined in IEEE 1609.2 and IEEE 1609.2.1. It then provides a detailed overview of the certificate request process. Finally, the section introduces the PQC algorithms that have been standardized by the NIST, as well as those currently under consideration for standardization.

### *A. The Architecture of SCMS and Data Structures*

This section begins by introducing the overall architecture of the SCMS, followed by an in-depth explanation of the certificate structure and the structure of the SPDU. Since the ETSI TS 103 097 standard defined by the ETSI considers only explicit certificates [6], and currently no implicit certificate schemes compliant with PQC standards are available [21], the discussion on certificate structure in this study focuses on explicit certificates.

#### 1). The Architecture of SCMS

The SCMS is primarily built upon the foundation of Public Key Infrastructure (PKI)[22] to provide secure certificates, as illustrated in **Fig. 1**.

In a SCMS, the trust anchor is established by the RCA. The certificates issued by the RCAs of certified certificate providers are included in the Certificate Trust List (CTL). As defined in IEEE 1609.2.1, the CTL is currently maintained and published by the SCMS Manager in the United States on its official website [7],[23]. Each certificate provider uses its RCA to issue certificates to its ICAs, which in turn issue certificates to the ECA, the PCA, and the RA. These communications are conducted in an I2I communications, which is not subject to the 1400-byte packet length limitation.

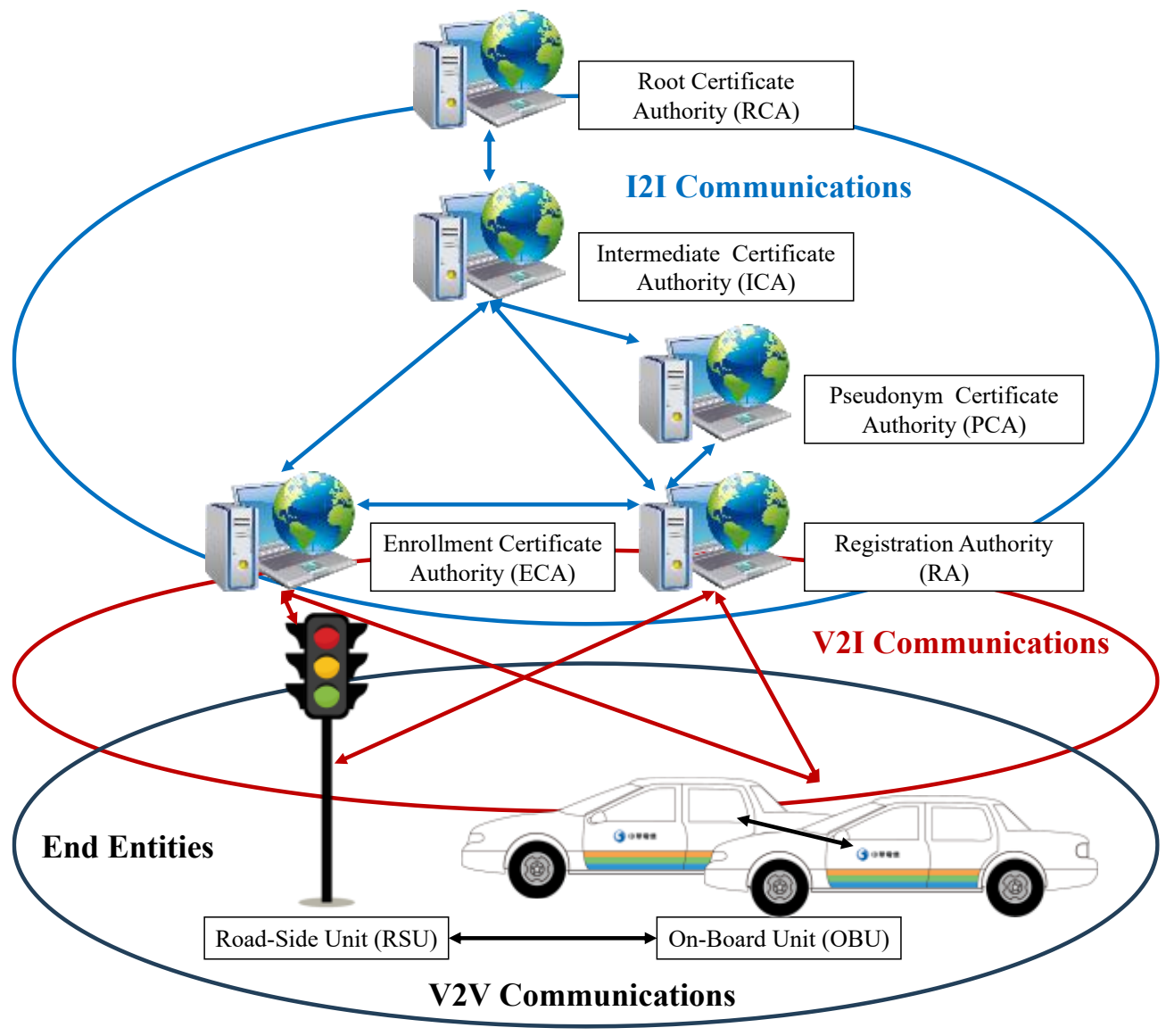

**Fig. 1.** The architecture of SCMS.

EEs are equipped with two types of certificates. The first is the EC, which functions as an identity credential verifying that the device is legitimately registered. This certificate is issued by the ECA and is used by the EE to request PCs. The second is the PC, which is required when the EE transmits Intelligent Transportation System (ITS) messages, such as BSMs. The PC serves to authenticate the sender as an authorized and trusted device. EEs request PCs from the PCA through the RA, and the PCA subsequently issues the PCs. Both the EC request and the PC request are carried out in V2I communications, and therefore are not constrained by the 1400-byte packet length limitation.

Once an EE has obtained a PC, it can participate in V2V communications to transmit ITS messages. For instance, OBUs may transmit BSMs, and RSUs may transmit Signal Phase and Timing (SPaT) messages. However, messages transmitted in V2V communications are subject to the 1400-byte packet length limitation defined by the IEEE 1609.3 standard [18]. As a result, strict control over message length is required to ensure compliance with this constraint.

#### 2). The Structure of Certificate

The certificate structure used in SCMS is applied consistently to both ECs and PCs, as illustrated in **Fig. 2**. The primary fields in the certificate include the Version, Type, Issuer, To-Be-Signed-Certificate (TBSC), and Signature [5],[24].

This study focuses exclusively on explicit certificates; therefore, the Type field is set to indicate an explicit certificate. The Issuer field contains the HashedId8 value of the CA that issued the certificate, serving as the issuer's identifier. For instance, the Issuer field in an EE’s EC corresponds to the HashedId8 of the ECA’s certificate, while the Issuer field in a pseudonym certificate corresponds to the HashedId8 of the PCA’s certificate. The TBSC component includes key certificate attributes such as the Validity Period, application permissions, and the Verify Key Indicator (VKI), which represents the public key used by the certificate holder for signature verification. Lastly, the Signature field contains the signature generated by the issuer using its private signing key to sign the TBSC. For example, the signature in an EC is generated by the ECA, while the signature in a PC is produced by the PCA.

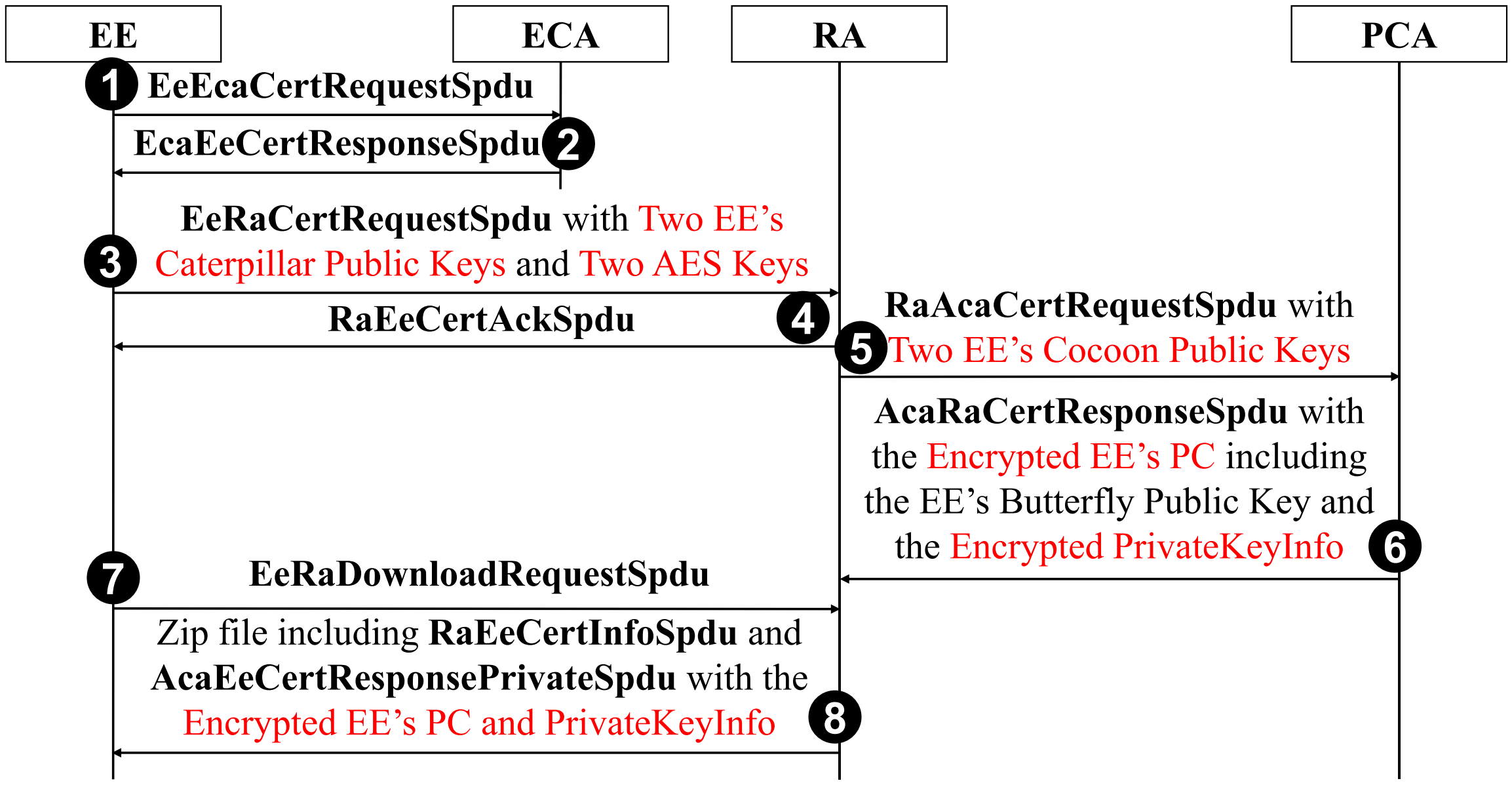

**Fig. 4.** The steps of EC and PC requests and responses.

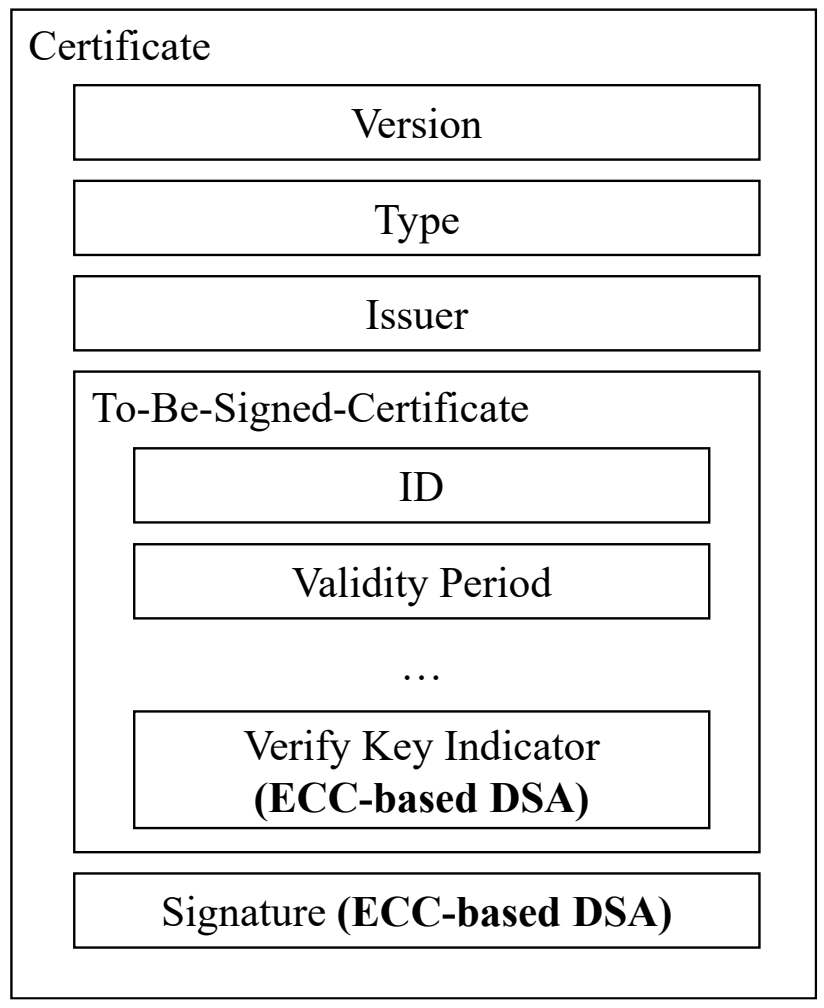

**Fig. 2.** The structure of certificate.

**3). The Structure of Signed SPDU**

The IEEE 1609.2 standard defines various types of SPDUs, including signed SPDUs and encrypted SPDUs [5]. The IEEE 1609.2.1 standard further introduces SPDUs for certificate requests and certificate responses [7]. This section primarily focuses on the structure of the signed SPDU, as shown in **Fig. 3**. A signed SPDU consists of Signed Data, which includes the To-Be-Signed-Data (TBSD), the Signer Identifier, and the Signature.

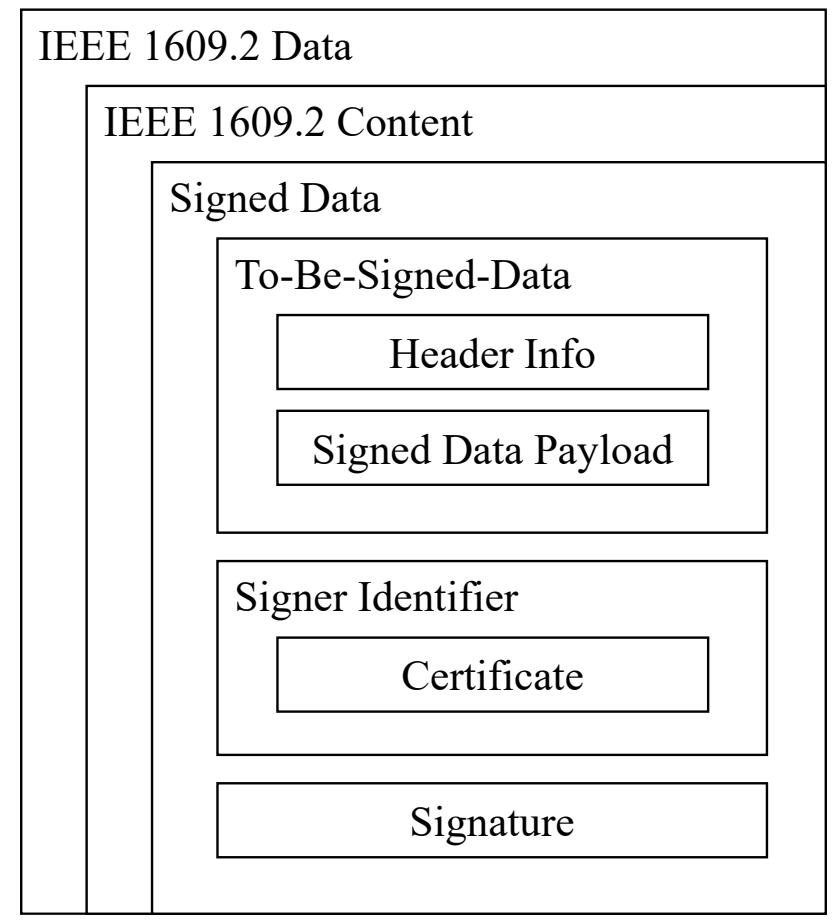

**Fig. 3.** The structure of signed SPDU.

For instance, in the case of an end entity transmitting a BSM, the TBSD field includes Header Information and the Signed Data Payload. The Header Information typically contains the Provider Service Identifier (PSID), the generation time, and other metadata. The Signed Data Payload carries the actual ITS message, which in this case is the BSM. The Signer Identifier field contains either the full certificate of the signer or a digest of that certificate [25]. If the full certificate is included, the total size of the SPDU increases accordingly. Finally, the Signature field holds the signature generated by the EE using its private key to sign the TBSD.

*B. The Steps of EC and PC Requests and Responses*

This section sequentially introduces the EC request process and the PC request process, as illustrated in **Fig. 4**. Following the description of these processes, the security of PCs is discussed. Lastly, the structure of certificate request and response SPDUs defined in IEEE 1609.2.1 is presented. This study primarily focuses on the Original BKE, while the Unified BKE and the Compact Unified BKE are considered as extensions derived from the Original BKE.

**1). The Steps of EC Request and Response**

During the manufacturing stage, each EE is pre-configured with a canonical key pair. The manufacturer registers the corresponding canonical identifier and canonical public key with the ECA. According to the IEEE 1609.2.1 standard, the canonical key pair is based on ECC.

When the EE is sold and enters operational use, it initiates the EC request process by sending an EeEcaCertRequestSpdu to the ECA. This message is a signed certificate request SPDU, as described in Section II.B.4. The EE generates an enrollment key pair and constructs a To-Be-Signed Request (TBSR) that includes the enrollment public key, application permissions, and other required information. The TBSR is then signed using the canonical private key through the Elliptic-Curve Digital Signature Algorithm (ECDSA). The resulting signature, along with the canonical identifier, is included in the EeEcaCertRequestSpdu and transmitted to the ECA (see **Fig. 4**❶). The primary computational overhead on the EE side lies in the ECC key generation and the ECDSA signature generation process.

Upon receiving the EeEcaCertRequestSpdu, the ECA retrieves the corresponding canonical public key based on the provided canonical identifier and verifies the included signature. Once the signature and the TBSR are successfully validated, the ECA issues an EC including the EE's enrollment public key for the EE. The issued certificate is embedded in the signed data of an EcaEeCertResponseSpdu, which is another type of signed SPDU. The ECA then generates an ECDSA signature over the signed data using its private key and includes it in the response message before sending it back to the EE (see **Fig. 4**❷). The main computational costs in this process are incurred during the verification of the EE's signature and the generation of two ECA's signatures for the EC and the SPDU.

**2). The Steps of PC Requests and Responses with BKE**

After obtaining the EC, the EE must request PCs. These PCs, along with the corresponding signatures, are required when transmitting ITS messages in order to authenticate that the messages originate from a trusted source. To ensure vehicular privacy, this process is explained using the original BKE mechanism, as described in prior studies.

Initially, the ee generates two ECC key pairs: one caterpillar key pair for signature and one caterpillar key pair for encryption. In addition, it generates two Advanced Encryption Standard (AES) keys to be used later in the Key Expansion Functions (KEFs). These components include the caterpillar public key for signature, the caterpillar public key for encryption, and the two AES keys. All are included in a TBSR, which is signed using the enrollment private key. The signed TBSR, along with the EC, is placed into a signed certificate request SPDU. This SPDU is then encrypted using the public key of the RA through the Elliptic-Curve Integrated Encryption Scheme (ECIES). The resulting encrypted SPDU (as described in Section II.B.4), referred to as EeRaCertRequestSpdu, is transmitted to the RA as shown in **Fig. 4**❸. The major computational efforts on the EE's side are attributed to the generation of the ECC key pairs and the ECIES encryption process.

Upon receiving the EeRaCertRequestSpdu, the RA decrypts the encrypted SPDU and extracts the enrollment public key from the EC to verify the signature contained within the request. Once the signature and the TBSR are confirmed to be valid, the RA generates a response called RaEeCertAckSpdu. This signed SPDU notifies the EE of the download schedule for the PCs, as shown in **Fig. 4**❹. The SPDU includes the RA's signature over the signed content. The primary computational efforts at this step involve ECIES decryption, verification of the EE's ECDSA signature, and generation of the RA's own ECDSA signature.

Next, the RA performs KEFs, as defined in IEEE 1609.2.1, based on the AES keys provided in the TBSR [7]. Thess KEFs, a type of Key Derivation Function (KDF), are applied to both the caterpillar public key for signature and the caterpillar public key for encryption in order to generate the corresponding cocoon public keys. These derived keys, referred to as the cocoon public key for signature and cocoon public key for encryption, are placed into a new signed TBSR. The RA then generates a signature over the new TBSR and includes it in a new signed certificate request SPDU, called RaAcaCertRequestSpdu, which is sent to the PCA, as shown in **Fig. 4**❺. The main computational requirements for this step include the derivation of the cocoon public keys, which is computationally similar to generating two ECC key pairs [26], and the production of the RA's ECDSA signature.

Upon receiving the RaAcaCertRequestSpdu, the PCA uses the RA's public key to verify the signature attached to the request. Once the signature and the contents of the TBSR are confirmed to be valid, the PCA generates a random number (referred to as PrivateKeyInfo). This value is then combined with the cocoon public key for signature included in the TBSR to derive the butterfly public key for signature through the BKE process. Using this derived key, the PCA issues the EE's PC, which includes the butterfly public key for signature. Subsequently, the PCA encrypts both the PC and the generated random number (i.e., PrivateKeyInfo) using the cocoon public key for encryption, in combination with the ECIES. The result is an encrypted SPDU. The PCA then signs the encrypted SPDU to produce a signature value and includes it in the AcaRaCertResponseSpdu, which is returned to the RA, as shown in **Fig. 4**❻. The primary computational costs at this step arise from verifying the RA's signature using the ECDSA, deriving the butterfly public key for signature (which is computationally equivalent to generating an ECC key pair [26]), generating the ECDSA signature for the PC, and generating the ECDSA signature for the AcaRaCertResponseSpdu.

When the scheduled time for downloading PC arrives, the EE constructs a signed SPDU using its enrollment private key to generate a signature. This signed SPDU is then encrypted with the RA's public key via ECIES to form the EeRaDownloadRequestSpdu, which is transmitted to the RA as depicted in **Fig. 4**❼. The main computational workload on the EE includes the generation of the ECDSA signature for the signed SPDU and the ECIES encryption operation.

Finally, the RA sends a compressed zip archive containing the ReEeCertInfoSpdu and the AcaEeCertResponsePrivateSpdu to the EE, as illustrated in **Fig. 4**❽. Since the RA's operations in this phase primarily involve database querying and file compression, which are not cryptographic in nature, these steps are not further analyzed in terms of cryptographic computational cost. The focus in this step lies on the cryptographic computations performed by the EE after receiving the compressed archive. The ReEeCertInfoSpdu only describes the timing of the next PC download and does not include a signature; therefore, no signature verification is required. In contrast, the AcaEeCertResponsePrivateSpdu contains a signature from the PCA and an encrypted payload. The EE first verifies the PCA's signature using the PCA's public key. Upon successful verification, the EE decrypts the payload using ECIES to obtain the PC and the corresponding random number (i.e., PrivateKeyInfo). Consequently, the primary cryptographic operations at this step for the EE include verifying the PCA's ECDSA signature and performing the ECIES decryption. The computations required to derive the cocoon private key for signature, cocoon private key for encryption, and butterfly private key for signature involve only simple integer additions, and thus their computational cost is negligible.

**3). Security Discussions of PC**

ECC is fundamentally based on the hardness of the Elliptic-Curve Discrete Logarithm Problem (ECDLP). Under current computational assumptions, classical computers are not capable of solving the ECDLP within polynomial time, as supported by prior studies [27],[28]. As a result, it is computationally infeasible to derive an ECC private key from its corresponding public key. The security analysis of the PC mechanism in this section is built on this assumption.

From the perspective of the PCA, the EE transmits its EC, which includes the enrollment public key and the caterpillar public key for signature, to the RA using encryption based on the RA's ECC public key. The RA then derives the cocoon public key for signature from the caterpillar public key for signature and sends it to the PCA. If the PCA cannot solve the ECDLP within polynomial time, it is unable to decrypt the ciphertext sent by the EE to the RA. It is also unable to determine the relationship between the cocoon public key for signature and the caterpillar public key for signature or establish any association between the enrollment public key and the cocoon public key for signature.

From the perspective of the RA, the PCA further expands the cocoon public key for signature using a random value, referred to as PrivateKeyInfo, to produce the butterfly public key for signature. This key is embedded in the PC. The PC and the associated PrivateKeyInfo are encrypted using the cocoon public key for encryption and processed with the ECIES. If the RA is not able to solve the ECDLP within polynomial time, it cannot decrypt the ciphertext received from the PCA. Additionally, it cannot infer the relationship between the cocoon public key and the butterfly public key for signature, nor can it deduce any relationship between the enrollment public key and the butterfly public key for signature.

From the perspective of other EEs, assuming they also cannot solve the ECDLP in polynomial time, it is infeasible for them to decrypt the ciphertext sent by the EE to the RA. It is equally infeasible to decrypt the ciphertext sent by the PCA to the EE or to derive any association between the enrollment public key and the butterfly public key for signature.

In summary, under the assumption that the ECDLP remains intractable within polynomial time for all involved parties, including the PCA, RA, and other EEs, it is not possible to uncover any relationship between the enrollment public key and the butterfly public key for signature. This property ensures the confidentiality and unlinkability of the PC system. However, since the system's security relies on the hardness of the ECDLP, it may become vulnerable in the presence of quantum computing technologies [12].

**4). The Structure of Signed Certificate Request SPDU**

The structure of the signed certificate request SPDU is defined in the IEEE 1609.2.1 standard [7], as illustrated in **Fig. 5**. This SPDU format includes a Signed Certificate Request, which consists of three main components: the TBSR, the Signer Identifier, and the Signature. In the case where the end entity transmits an EeEcaCertRequestSpdu, the "Request" component shown in Fig. 5 refers specifically to the EeEcaCertRequest.

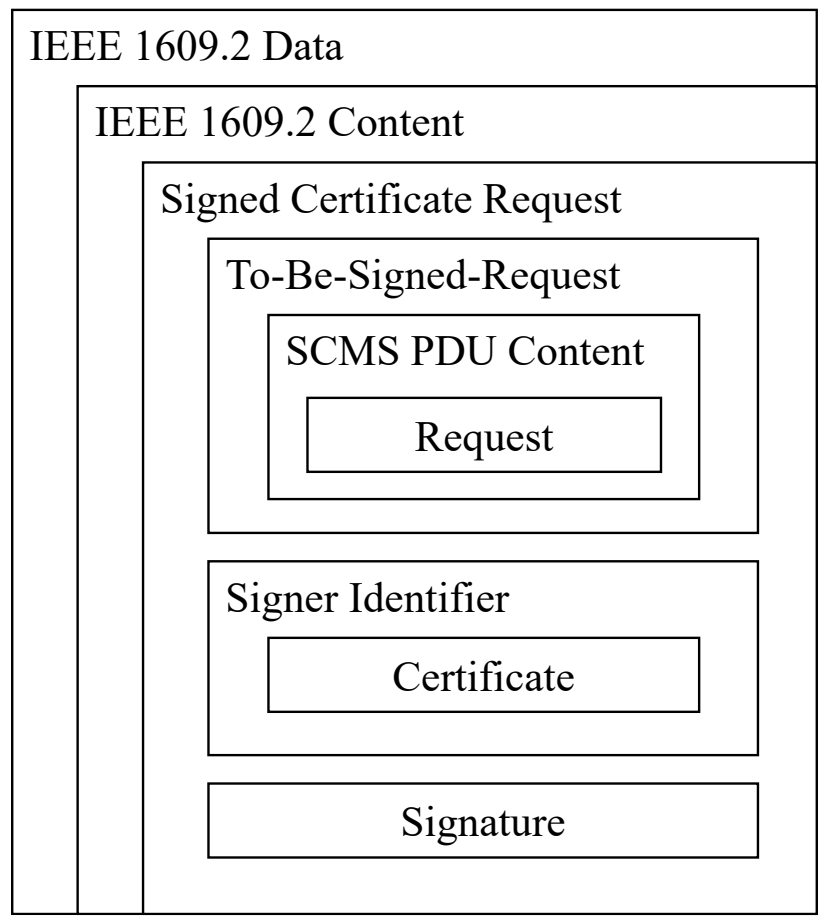

**Fig. 5.** The structure of signed certificate request SPDU.

The structure of the encrypted SPDU is defined in the IEEE 1609.2 standard [5], as illustrated in **Fig. 6**. The format of the encrypted SPDU includes an Encrypted Data field, which contains the Recipients and the Ciphertext. According to IEEE 1609.2.1 [7], the Recipients field can include PreSharedKeyRecipientInfo, SymmRecipientInfo, or PKRecipientInfo. As the PC process defined in IEEE 1609.2.1 primarily employs an ECIES [29], which uses the recipient's ECC public key for encryption, this section focuses on PKRecipientInfo.

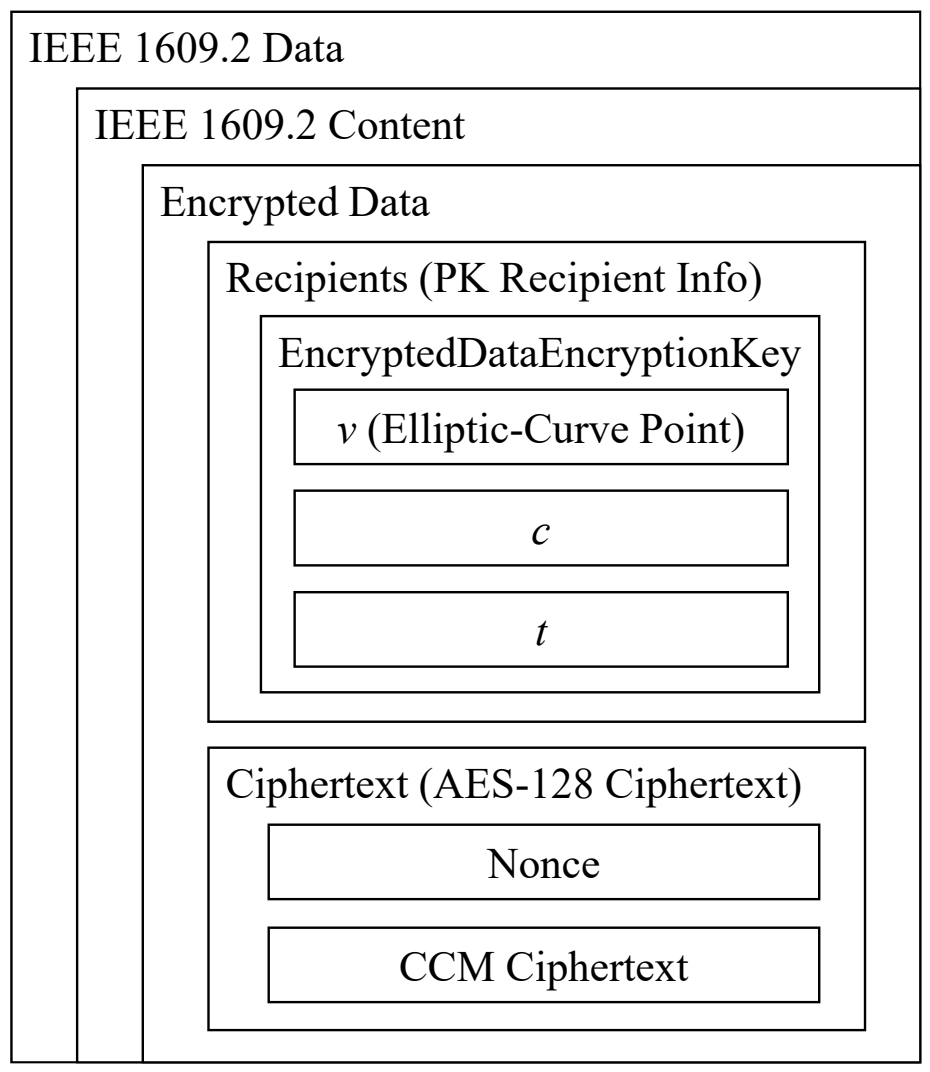

**Fig. 6.** The structure of encrypted SPDU.

In the context of this study, the Key-Encryption-Key (KEK) is assumed to be a 256-bit random value, and the Data-Encryption-Key (DEK) is assumed to be a 128-bit random value used for AES-128 encryption. During the ECIES process, a random value is generated dynamically and used to derive an elliptic-curve point, represented as parameter $v$ in **Fig. 6**. The Elliptic-Curve Diffie–Hellman (ECDH) algorithm is then used to derive a shared secret. If the elliptic-curve used is NIST P-256, the resulting shared secret is a 256-bit value that serves as the KEK. A separate 128-bit random value is then generated as the DEK. Using the key wrapping mechanism specified in the IEEE 1363a standard [30], the DEK is wrapped using the KEK. This produces a wrapped ciphertext, denoted as parameter $c$ in **Fig. 6**, along with a keyed-Hash Message Authentication Code (HMAC) generated using the Secure Hash Algorithm 256 (SHA-256), denoted as parameter $t$ in **Fig. 6**. The Ciphertext field primarily stores the encrypted data obtained using the DEK.

### *C. PQC Standards*

This section focuses on PQC algorithms that have been standardized or are under consideration for standardization by the NIST. Section II.C.1 presents KEMs, followed by an overview of Digital Signature Algorithms (DSAs) in Section II.C.2.

#### 1). Key Encapsulation Mechanisms

Based on evaluation criteria such as resistance to quantum attacks, key pair generation time, key encapsulation time, and key decapsulation time, the NIST selected suitable KEMs. In the report of the 3rd PQC standardization conference released in 2022, the Cryptographic Suite for Algebraic Lattices (CRYSTALS)-Kyber was identified as a candidate for standardization [31]. The algorithm was officially standardized in 2024 under the name ML-KEM (formerly known as CRYSTALS-Kyber) [13]. This algorithm is characterized by relatively short key pair generation, encapsulation, and decapsulation times.

Furthermore, the report of the 4th PQC standardization conference, published in 2025, selected another algorithm, HQC (Hamming Quasi-Cyclic), as a candidate for standardization [32]. Although McEliece was not selected by NIST as a standardization candidate, the report noted that it is under consideration for standardization by the International Organization for Standardization (ISO) [32]. Both HQC and McEliece are based on code-based cryptography, which differs from lattice-based cryptography in its underlying hard problems. McEliece offers shorter encapsulation and decapsulation times but requires more time for key pair generation, which contributed to its exclusion from NIST's list of standardization candidates.

#### 2). Digital Signature Algorithms

The NIST has evaluated DSAs based on several criteria, including resistance to quantum attacks, key pair generation time, signature generation time, signature verification time, public key length, and signature length. In the report of the 3rd PQC standardization conference released in 2022, three algorithms were selected as candidates for standardization: CRYSTALS-Dilithium, SPHINCS+, and Falcon [31]. Among them, the ML-DSA (formerly CRYSTALS-Dilithium) [14] and the SLH-DSA (formerly SPHINCS+) [15] were officially standardized in 2024. The ML-DSA is characterized by relatively low key pair generation time, signature generation time, and verification time. The SLH-DSA offers advantages such as short private and public key lengths, with its security grounded in well-established hash functions. Falcon, although not yet standardized, remains under evaluation due to its benefits, including shorter signature lengths and faster signing performance.

While these three algorithms have been selected for standardization, the signature lengths are still considered relatively large. In response, NIST has initiated an evaluation process aimed at identifying additional digital signature schemes that offer shorter signature lengths and lower computational costs. In 2024, the 1st report on this additional digital signature scheme evaluation was published, identifying 14 preliminary candidates. However, no final selection has been made thus far [33]. Given this context, the present study focuses on comparing the standardized or selected algorithms: the ML-DSA, the SLH-DSA, and Falcon.

## III. The Proposed Pseudonym Scheme Based on Hybrid Certificates

This section presents the proposed certificate structure in this study and its applicable scenarios in Section III.A. The proposed general pseudonym scheme and its security properties are illustrated in Section III.B.

### *A. The Proposed Certificate Structure and Its Applicable Scenarios*

Given that there is no 1400-byte packet length restriction in I2I and V2I communications, whereas V2V communications are subject to this constraint, a pure PQC certificate (as introduced in Section III.A.1) is not suitable for V2V applications. In response, this study proposes a hybrid certificate (as described in Section III.A.2), which is designed to be applicable to V2V communication while still offering enhanced protection for infrastructure devices including RCA, ICA, ECA, PCA, and RA. Section III.A.3 provides a comparative analysis of the security properties of both the pure PQC and hybrid certificate schemes.

**1). The Structure of Pure PQC Certificate for Infrastructure Device's Certificate and EE's EC**

To enhance resilience against quantum computing threats, this study builds upon the IEEE 1609.2 standard and proposes a pure PQC certificate structure, illustrated in **Fig. 7**. In this structure, the Signature field contains a signature generated using a PQC-based DSA, issued by a CA employing PQC. Meanwhile, the VKI field carries a public key generated using a PQC-based DSA, which corresponds to the certificate holder's cryptographic identity.

The PQC-based DSAs referenced in this section include ML-DSA, SLH-DSA, and Falcon. Should additional PQC algorithms be standardized in the future, they can also be adopted within this structure. Importantly, the algorithm used by the CA to generate the signature and the one used by the certificate holder to create their public key may differ.

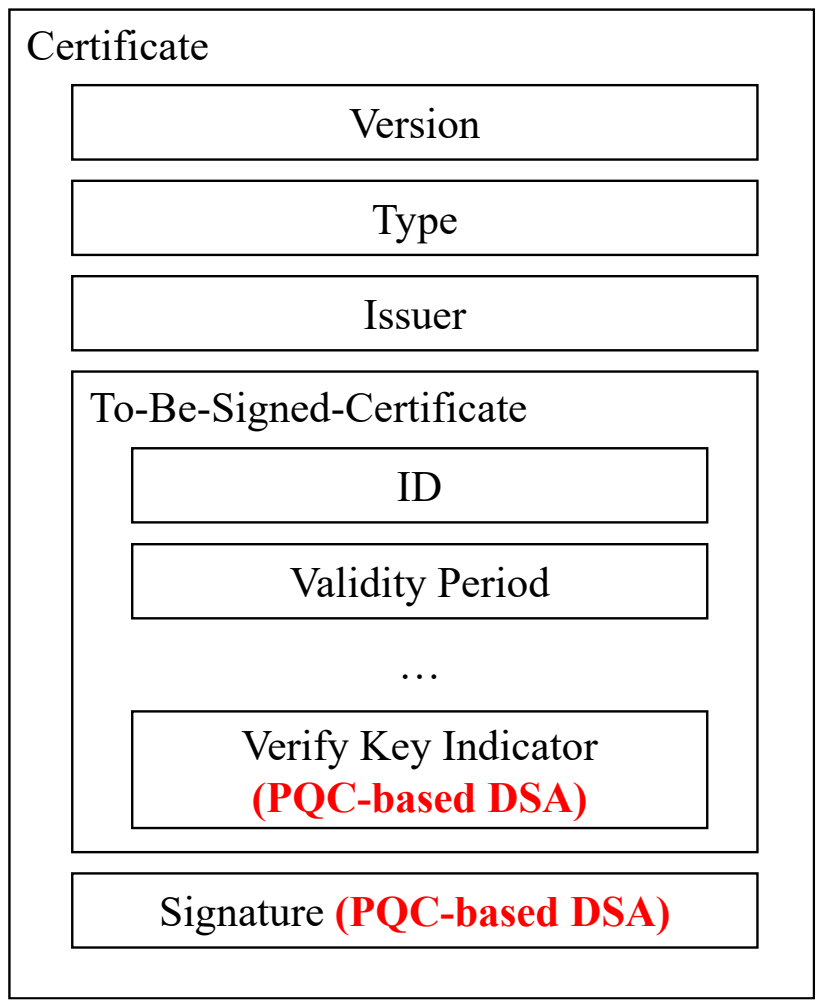

**Fig. 7.** The structure of pure PQC certificate.

A key advantage of the pure PQC certificate lies in its end-to-end post-quantum security: both the CA and the certificate holder utilize cryptographic methods believed to be secure against quantum adversaries. However, a notable drawback is the increased certificate size. For instance, ML-DSA-44 produces a signature of approximately 2420 bytes. When such algorithms are used, the certificate size exceeds the 1400-byte limit imposed on V2V packets. At present, there is no standardized PQC method capable of producing a pure PQC certificate within the 1400-byte constraint. Consequently, the use of pure PQC certificates is limited to I2I and V2I communication scenarios. This makes them suitable for interactions among the RCA, ICA, ECA, PCA, and RA, all of which typically operate over I2I links. Additionally, since ECs for EEs are only transmitted in V2I contexts and not in V2V communications, they can also adopt the pure PQC certificate format to maximize security.

**2). The Structure of Hybrid Certificate for EE's PC**

To address the need for both quantum-resistance and compatibility with V2V communication, this study proposes a hybrid certificate structure based on the IEEE 1609.2 standard, as illustrated in **Fig. 8**. In this design, the Signature field contains a signature generated using a PQC-based DSA, indicating that the CA employs a PQC-based scheme. Simultaneously, the VKI field carries a public key generated using an ECC-based DSA, meaning that the certificate holder (i.e., the EE) utilizes ECC-based DSA.

The ECC-based DSAs referenced in this context may include ECDSA, SM2, or may be substituted with RSA-based DSAs. Importantly, the CA continues to utilize a PQC-based DSA, thereby maintaining quantum-resistance for infrastructure-level entities.

The main advantage of the hybrid certificate lies in its ability to combine the quantum-resistance of PQC (used by the CA) with the compactness of ECC-based public keys (used by the EE). This structure allows the overall certificate size to remain within the 1400-byte packet size constraint required by V2V communication. For instance, Falcon-512 produces a signature of 666 bytes, and an ECDSA NIST P-256 public key (e.g., a compressed elliptic-curve point) is approximately 33 bytes, enabling the hybrid certificate to meet the size requirements for V2V scenarios such as EE's PCs. However, a key limitation of the hybrid certificate is that the use of ECC-based DSA by EEs may be vulnerable to compromise by quantum adversaries, as such schemes can potentially be broken in polynomial time using quantum algorithms.

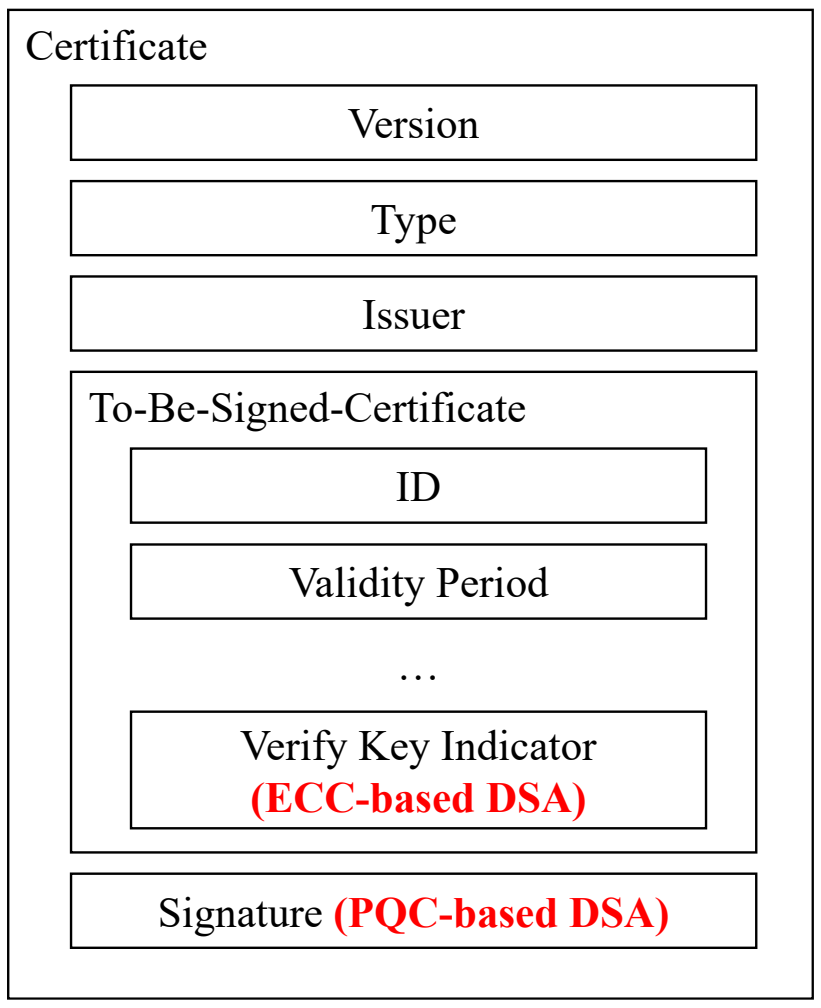

**Fig. 8.** The structure of hybrid certificate.

**3). The Security of the Proposed Certificates**

The PQC-based DSAs referenced in this study refer to those officially selected and standardized by the NIST. These algorithms have been theoretically validated by the international cryptographic community to possess resistance against quantum computing attacks. Each algorithm is based on distinct computationally hard problems, with technical details documented in the reports of the third and fourth NIST PQC standardization conferences [31], [32]. The use of PQC-based DSAs in this study is not restricted to a single implementation. The cryptographic design remains adaptable, allowing for flexible substitution of algorithms. This adaptability ensures that, should any vulnerabilities be identified in a standardized algorithm in the future, it is feasible to transition to another secure alternative within the same certificate structure. As a result, certificates based entirely on PQC signatures retain the capability to resist quantum attacks.

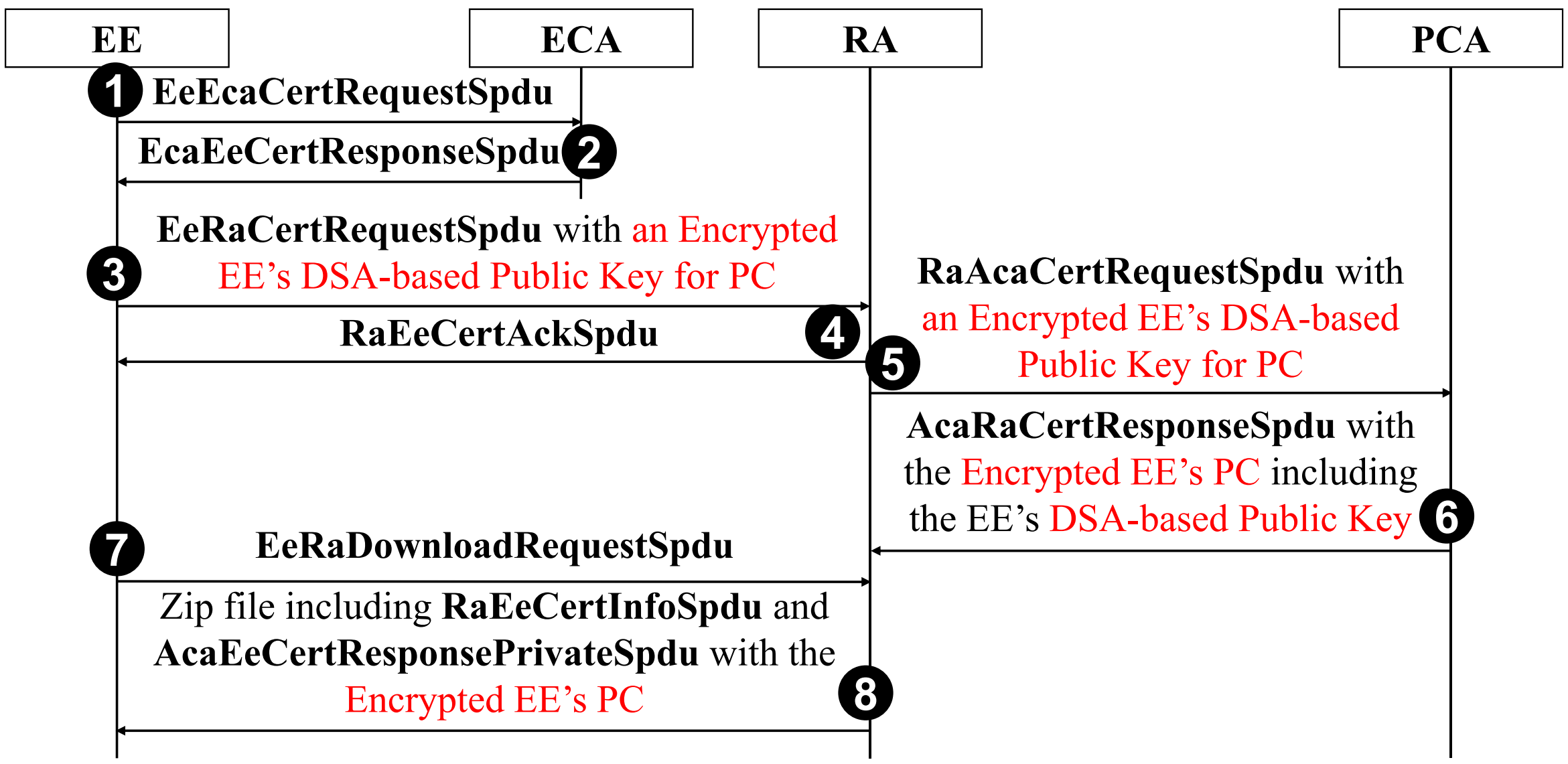

**Fig. 9.** The steps of the proposed pseudonym scheme based on hybrid certificates.

While hybrid certificates are suitable for V2V communication and PCs, they rely on ECC signatures on the part of the certificate holder (i.e., EEs). As ECC-based DSAs are potentially vulnerable to quantum attacks, this introduces a risk. To mitigate such risks, it is recommended that PCs be assigned shorter validity periods. For instance, if quantum computing were to require approximately one week to break the ECDLP under the NIST P-256 parameters, then a pseudonym certificate validity period of one day would ensure that the certificate expires before it can be compromised. In the long term, full migration to pure PQC certificates will depend on the availability of suitable standardized DSAs that satisfy all operational constraints. In the interim, hybrid certificates represent a practical and immediately deployable transitional solution.

### *B. The Proposed Generalized Pseudonym Scheme*

The proposed generalized pseudonym scheme and its detailed procedure are illustrated in **Fig. 9** and described in Section III.B.1. Section III.B.2 provides a discussion of the scheme's security, followed by Section III.B.3, which presents the structure of the proposed KEM-based encrypted SPDU.

#### 1). The Steps of the Proposed Scheme

During the manufacturing phase, each EE is pre-configured with a PQC-based canonical key pair, and the corresponding canonical identifier and PQC-based canonical public key are registered with an ECA, thereby enabling future certificate requests to support quantum-resistant security.

Once deployed and operational, the EE generates a PQC-based enrollment key pair. The device then uses its PQC-based canonical private key to sign a certificate request, referred to as a TBSR, which includes the PQC-based enrollment public key and the associated application permissions. This signature, along with the canonical identifier, is included in the EeEcaCertRequestSpdu and transmitted to the ECA, as illustrated in **Fig. 9** ❶. The primary computational tasks for the device include the generation of the PQC-based enrollment key pair and the generation of the PQC-based signature.

Upon receiving the EeEcaCertRequestSpdu, the ECA retrieves the associated PQC-based canonical public key based on the canonical identifier. It then verifies the signature contained in the request using this public key. If the signature and the request contents are validated successfully, the ECA issues an EC for the EE, which contains the PQC-based enrollment public key. The RCA includes this certificate in the TBSD of the EcaEeCertResponseSpdu, signs the data using its private key with a PQC-based DSA, and returns the signed message to the EE, as shown in **Fig. 9**❷. On the ECA side, the main computational overhead arises from verifying the PQC-based signature submitted by the EE and generating two PQC-based signatures for the EC and the SPDU.

Following the issuance of the EC, the EE initiates the PC request process. As part of this procedure, the EE generates a DSA-based key pair. The DSA-based public key from this pair is then encrypted using the public key of the PQC, which is constructed using a PQC-based KEM. The technical aspects of this encryption process are described in Section III.B.3. The resulting ciphertext is embedded within the TBSR. The EE signs this request using its PQC-based enrollment private key. Both the signed request and the EC are then packaged into a signed certificate request SPDU for the certificate request. Subsequently, this SPDU is encrypted using the RA's PQC-based KEM public key to produce the encrypted SPDU, referred to as EeRaCertRequestSpdu. The EeRaCertRequestSpdu is transmitted to the RA, as illustrated in **Fig. 9**❸. The primary computational efforts required in this process include generating the DSA-based key pair, producing the signature on the TBSR using a PQC-based DSA, and performing two PQC-based KEM encapsulation operations for encryption. The first encapsulation is performed with the PCA's public key, and the second with the RA's public key.

Upon receiving the EeRaCertRequestSpdu, the RA uses its PQC-based KEM private key to decrypt the EeRaCertRequestSpdu, and the PQC-based enrollment public key is extracted from the included EC to verify the signature attached to the request. Once the signature and the contents of the TBSR are successfully validated, the RA responds with an RaEeCertAckSpdu, which serves as a signed SPDU. This response notifies the EE of the time at which the PC will be available for download, as illustrated in Fig. 9❹. The RaEeCertAckSpdu includes a signature generated by the RA over the corresponding signed data. The primary computational workload in this process involves decrypting the received data using the PQC-based KEM private key, verifying the PQC-based signature provided by the EE, and generating the RA's own PQC-based signature.

Subsequently, the RA includes the ciphertext of the DSA-based public key, originally contained in the TBSR within the EeRaCertRequestSpdu, into the TBSR of the RaAcaCertRequestSpdu. This request is then combined with a signature generated by the RA and sent to the PCA, as illustrated in **Fig. 9**❺. The primary computational workload at this step involves generating the RA's signature using a PQC-based DSA.

Upon receiving the RaAcaCertRequestSpdu, the PCA verifies the signature using the RA's PQC-based public key. Once the validity of the signature and the integrity of the TBSR are confirmed, the PCA uses its PQC-based KEM private key to decrypt the ciphertext containing the DSA-based public key from the TBSR. A PC for the EE is then issued, embedding the plaintext DSA-based public key within the PC. Since the decryption process also yields a pre-shared key, the PC is subsequently encrypted using this pre-shared key to produce an encrypted SPDU. A signature is then generated over this encrypted SPDU and included in the AcaRaCertResponseSpdu, which is returned to the RA, as illustrated in **Fig. 9**❻. The primary computational workload in this process involves verifying the RA's signature using a PQC-based DSA, performing decryption with the PQC-based KEM private key, generating the signature embedded in the PC, and producing the signature included in the AcaRaCertResponseSpdu.

When the designated time for downloading the PC arrives, the EE generates a signed SPDU. This signature is created using the EE's PQC-based enrollment private key. The resulting signed SPDU is then encrypted using the RA's PQC-based KEM public key, producing an encrypted SPDU referred to as the EeRaDownloadRequestSpdu. This encrypted SPDU is transmitted to the RA, as illustrated in Fig. 9❼. The primary computational overhead on the EE arises from generating the signature within the SPDU using a PQC-based DSA, as well as performing encapsulation and encryption with the PQC-based KEM public key.

Finally, the RA transmits a compressed zip file containing both the ReEeCertInfoSpdu and the AcaEeCertResponsePrivateSpdu to the EE, as shown in **Fig. 9** ❽. On the side of the RA, the operations primarily involve data retrieval and compression, which are not cryptographic in nature and therefore are not discussed in detail here. The focus at this step is on the cryptographic computation time required by the EE upon receiving the compressed file. The ReEeCertInfoSpdu contains information about the next PC download time and does not include any signature, thereby eliminating the need for signature verification. In contrast, the AcaEeCertResponsePrivateSpdu includes a signature and ciphertext issued by the PCA. The EE first uses the PCA's PQC-based public key to verify the signature. Upon successful verification, the EE uses the pre-shared key to decrypt the ciphertext and retrieve the PC. Therefore, the primary computation time at the EE is spent on verifying the PQC-based signature and performing decryption using the pre-shared key.

**2). The Security of the Proposed Scheme**

The generalized pseudonym scheme proposed in this study adopts PQC techniques throughout both the certificate request and response procedures. It is designed to be compatible with any algorithm approved as a standard by the NIST, with its security based on the underlying cryptographic method employed [31],[32]. Regarding the use of PQC-based KEM, the process begins with the encapsulation of a KEK using a PQC-based KEM public key. This KEK is then used to encrypt a DEK, which in turn encrypts the relevant data. Therefore, the security of the KEK depends on the strength of the PQC-based KEM, while the DEK and the encrypted data rely on the security of the AES algorithm. Since current quantum computing capabilities do not allow for the compromise of AES within polynomial time, AES remains considered resistant to quantum attacks [31].

From the perspective of the PCA, the EE transmits ciphertext to the RA. This ciphertext includes the enrollment public key and the encrypted DSA-based public key intended to be placed within the PC. Encryption is performed using RA's PQC-based KEM public key. As long as the PCA cannot break either the PQC-based KEM or AES encryption in polynomial time, it will be unable to decrypt the ciphertext or determine any relationship between the enrollment public key and the DSA-based public key stored in the EE's PC.

From the perspective of the RA, the EE uses the PCA's PQC-based KEM public key to encrypt the DSA-based public key that will appear in the PC. The pc, after being issued by the PCA, is then encrypted using a pre-shared key and the AES algorithm before being forwarded to the RA. If the RA cannot break either the PQC-based KEM or AES encryption within polynomial time, it is similarly unable to decrypt the ciphertext transmitted by the EE to the PCA or to determine the association between the enrollment public key and the DSA-based public key in the EE's PC.

From the perspective of other EEs, if neither the PQC-based KEM nor the AES encryption can be broken in polynomial time, then these EEs will also be unable to decrypt any of the ciphertexts sent between the EE and the RA or between the EE and the PCA. As a result, they cannot deduce the relationship between the enrollment public key and the DSA-based public key contained in the EE's PC.

In conclusion, none of the involved parties, including the PCA, the RA, or any external EEs, can infer the connection between the enrollment public key and the DSA-based public key contained in the EE's PC. This guarantees the security of the PCs in terms of unlinkability. Additionally, the EE has the capability to request multiple PCs in a single session and to rotate them periodically to enhance privacy protection.

**3). The Structure of KEM-based Encrypted SPDU**

Based on the IEEE 1609.2 standard [5], this study designs a KEM-based encrypted SPDU, as illustrated in **Fig. 10** and **Fig. 11.**

**Fig. 10** represents the structure used by the requester during the request transmission. In this process, the requester encapsulates a KEK using the recipient's KEM-based public key. The resulting encapsulated key (i.e., the ciphertext of the KEK) is stored in the parameter $v$ as shown in **Fig. 10**. The KEK is then used to wrap a DEK using the key wrapping mechanism defined in the IEEE 1363a standard [30]. This process produces the wrapped ciphertext (represented as parameter $c$ in **Fig. 10**) and a HMAC generated using the SHA-256, denoted as parameter $t$. The encrypted data, obtained by encrypting the payload with the DEK using the AES CCM mode, is stored in the CCM Ciphertext field, and the associated nonce is placed in the Nonce field.

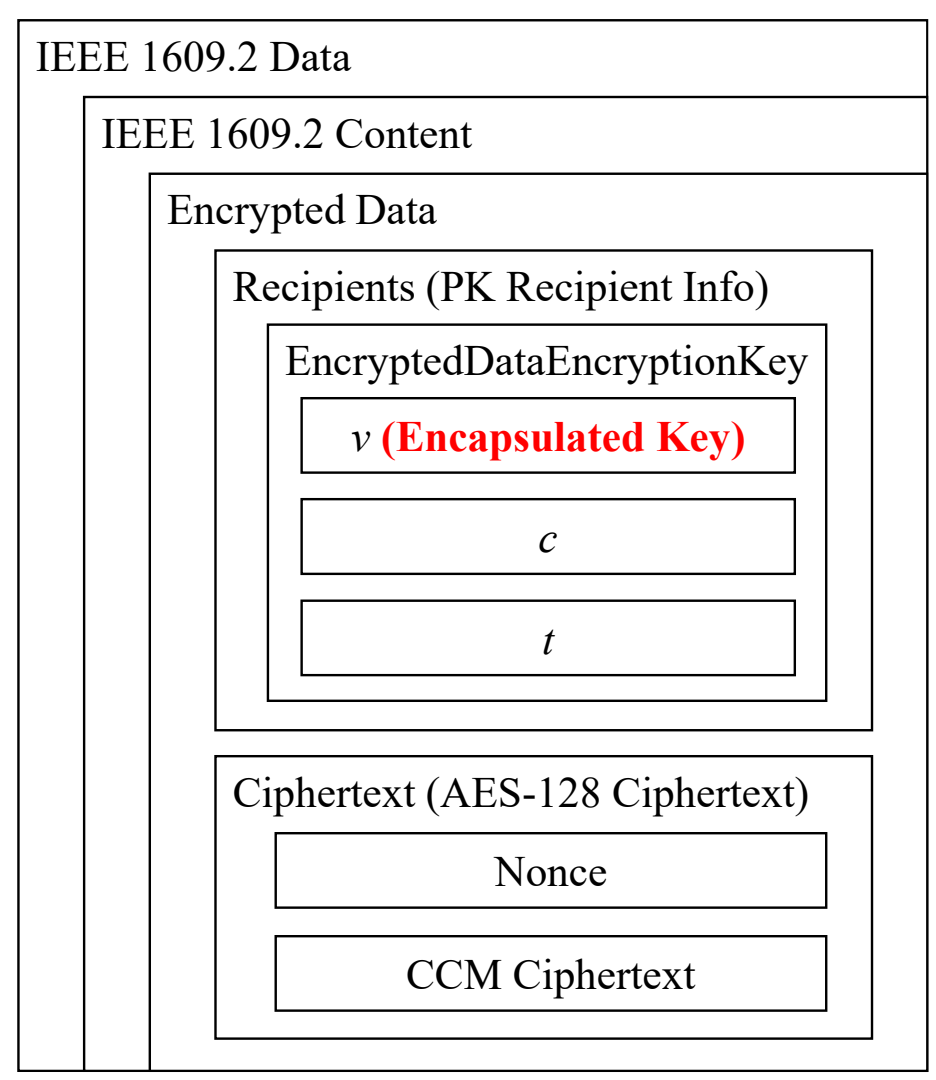

**Fig. 10.** The structure of KEM-based encrypted SPDU for requests.

**Fig. 11.** illustrates the structure used for the response. The recipient decrypts the KEK and unwraps the DEK, which is then used to encrypt the response data using AES CCM mode. The resulting ciphertext is stored in the CCM Ciphertext field, and the associated nonce is stored in the Nonce field. Since the DEK can serve as a pre-shared key, the Recipients field in the response structure is designated as PreSharedKeyRecipientInfo. This eliminates the need to encrypt or transmit the DEK ciphertext again using the recipient's public key.

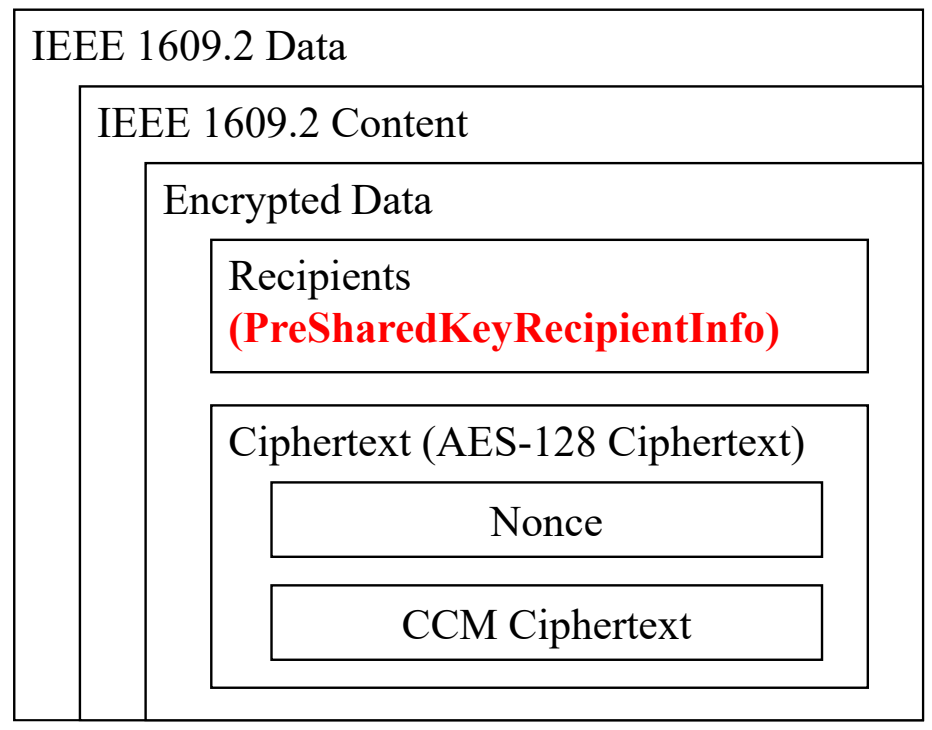

**Fig. 11.** The structure of KEM-based encrypted SPDU for responses.

As a result, the response format of the KEM-based encrypted SPDU proposed in this study achieves a shorter message length while maintaining cryptographic security.

## IV. Experimental Results and Discussions

This section presents the implementation and evaluation of the hybrid certificate-based pseudonym scheme proposed in this study, along with a performance comparison. Section IV.A describes the experimental environment and field setup. Sections IV.B and IV.C analyze and compare the certificate lengths and message lengths, respectively, while Section IV.D evaluates the computation time required to generate and verify each message. Furthermore, Section IV.E discusses the security of the proposed scheme and the protection of vehicular privacy. Finally, the migration issues and costs are illustrated in Section IV.F.

### *A. Experimental Environments*

Given that OBUs are typically embedded systems with constrained computational resources, Raspberry Pi 4 was selected as the EE for the evaluation in this study. Although some cryptographic processing is performed by the RA and CA, only the computational results from Raspberry Pi 4 are reported. These results can be considered a conservative baseline, as RA and CA components would typically be deployed on cloud servers with higher processing capabilities. The certificate of RCA developed in this study has successfully passed interoperability testing at the OmniAir Plugfest event and has been included in the Test CTL published by the SCMS Manager. Building on this foundation, the proposed hybrid certificate-based pseudonym scheme was integrated into the system and validated using real-world data.

For security level evaluation, the study adopts the definitions provided by the NIST for PQC algorithms. As RSA and ECC lack resistance against quantum computing attacks, they are not assigned any security level under the NIST definitions. Accordingly, this study designates such schemes as "Security Level 0" to represent non-quantum-safe configurations.

*B. The Comparison of Certificate Lengths*

This section analyzes the impact of various combinations of DSAs used by the CA and the certificate holder on the resulting certificate length, as well as their associated SLs, as summarized in **Table I**. The evaluation includes combinations of ECDSA, RSA, ML-DSA, Falcon, and SLH-DSA algorithms. Specifically, the study examines configurations in which the CA's DSA (corresponding to the certificate signature (Sig.)) and the certificate holder's DSA (corresponding to the public key in the certificate, i.e., VKI) are varied. The configurations include pure ECDSA, pure RSA, pure ML-DSA, pure Falcon, pure SLH-DSA, and hybrid combinations such as ML-DSA with ECDSA, ML-DSA with RSA, Falcon with ECDSA, Falcon with RSA, SLH-DSA with ECDSA, and SLH-DSA with RSA. In the table, hybrid certificates are highlighted with a green background, and configurations in which the combined signature and public key sizes are less than 1400 bytes are indicated in bold.

TABLE I.
THE COMPARISON OF CERTIFICATE LENGTHS (UNIT: BYTES) AND SECURITY LEVELS (0: UNSECURE; 5: HIGH SECURE)

| CA's DSA | Certificate Holder's DSA | Sig. | VKI | SL (Sig.) | SL (VKI) |
|---|---|---|---|---|---|
| **ECDSA NIST P-256** | **ECDSA P-256** | **65** | **33** | **0** | **0** |
| **ECDSA NIST P-384** | **ECDSA P-384** | **97** | **49** | **0** | **0** |
| **ECDSA NIST P-521** | **ECDSA P-521** | **132** | **67** | **0** | **0** |
| **RSA-3072** | **RSA-3072** | **384** | **422** | **0** | **0** |
| RSA-7680 | RSA-7680 | 960 | 998 | 0 | 0 |
| RSA-15360 | RSA-15360 | 1920 | 1958 | 0 | 0 |
| ML-DSA-44 | ML-DSA-44 | 2420 | 1312 | 2 | 2 |
| ML-DSA-65 | ML-DSA-65 | 3309 | 1952 | 3 | 3 |
| ML-DSA-87 | ML-DSA-87 | 4627 | 2592 | 5 | 5 |
| ML-DSA-44 | ECDSA NIST P-256 | 2420 | 33 | 2 | 0 |
| ML-DSA-65 | ECDSA NIST P-384 | 3309 | 49 | 3 | 0 |
| ML-DSA-87 | ECDSA NIST P-521 | 4627 | 67 | 5 | 0 |
| ML-DSA-44 | RSA-3072 | 2420 | 422 | 2 | 0 |
| ML-DSA-65 | RSA-7680 | 3309 | 998 | 3 | 0 |
| ML-DSA-87 | RSA-15360 | 4627 | 1958 | 5 | 0 |
| Falcon-512 | Falcon-512 | 666 | 897 | 1 | 1 |
| Falcon-1024 | Falcon-1024 | 1280 | 1793 | 5 | 5 |
| **Falcon-512** | **ECDSA NIST P-256** | **666** | **33** | **1** | **0** |
| **Falcon-1024** | **ECDSA NIST P-521** | **1280** | **67** | **5** | **0** |
| **Falcon-512** | **RSA-3072** | **666** | **422** | **1** | **0** |
| Falcon-1024 | RSA-15360 | 1280 | 1958 | 5 | 0 |
| SLH-DSA-SHA2-128s | SLH-DSA-SHA2-128s | 7856 | 32 | 1 | 1 |
| SLH-DSA-SHA2-192s | SLH-DSA-SHA2-192s | 16224 | 48 | 3 | 3 |
| SLH-DSA-SHA2-256s | SLH-DSA-SHA2-256s | 29792 | 64 | 5 | 5 |
| SLH-DSA-SHA2-128s | ECDSA NIST P-256 | 7856 | 33 | 1 | 0 |
| SLH-DSA-SHA2-192s | ECDSA NIST P-384 | 16224 | 49 | 3 | 0 |
| SLH-DSA-SHA2-256s | ECDSA NIST P-521 | 29792 | 67 | 5 | 0 |
| SLH-DSA-SHA2-128s | RSA-3072 | 7856 | 422 | 1 | 0 |
| SLH-DSA-SHA2-192s | RSA-7680 | 16224 | 998 | 3 | 0 |
| SLH-DSA-SHA2-256s | RSA-15360 | 29792 | 1958 | 5 | 0 |

The experimental results show that all pure ECDSA combinations satisfy the size constraints required for V2V communication. In the case of pure RSA, only configurations using 3072-bit keys meet the requirement. On the other hand, all combinations using pure PQC result in total sizes greater than 1400 bytes. As a result, they are unsuitable for V2V communications but may still be considered for I2I or V2I communications, where they can help protect infrastructure devices against quantum-based threats.

Among the hybrid certificate types, only those involving Falcon have the potential to meet the 1400-byte size constraint. Although the combination of Falcon-1024 and ECDSA NIST P-521 offers a higher level of security, the resulting signature and public key sizes are close to the 1400-byte limit, leaving insufficient space for additional certificate information. Therefore, this configuration is also considered impractical for V2V communication. In addition, since the signed SPDU with ITS message must include both the EE's PC and the signature, the combination of the Falcon-512 and RSA-3072 hybrid certificate (at least 666 + 422 = 1088 bytes) with an RSA-3072 signature (384 bytes) would exceed the 1400-byte limit. Therefore, only the hybrid certificate combining Falcon-512 and ECDSA NIST P-256 (marked by red color) remains as the sole viable option currently suitable for V2V communication.

Based on these findings, the remainder of this study assumes that Falcon-512 is used as the DSA by the ECA, the PCA, and the RA. In addition, Falcon-512 is also assumed for both the canonical key pair and enrollment key pair of the EE.

*C. The Comparison of Message Lengths*

This section presents a comparative analysis of the message lengths involved in EC and PC requests between the IEEE 1609 series standards and the hybrid certificate-based pseudonym scheme proposed in this study. Specifically, eight types of messages are examined in detail: EeEcaCertRequestSpdu, EcaEeCertResponseSpdu, EeRaCertRequestSpdu, RaEeCertAckSpdu, RaAcaCertRequestSpdu, AcaRaCertResponseSpdu, EeRaDownloadRequestSpdu, and AcaEeCertResponsePrivateSpdu, as shown in **Table II** and **Table III**. In addition, the study analyzes the practical application of PCs in transmitting ITS messages by generating signed SPDUs containing BSMs and comparing the resulting message lengths.

This study examines a typical EE responsible for transmitting BSMs, representing standard OBU functionality. It evaluates the complete process from the EE's EC and PC requests to the transmission of a signed SPDU containing the BSM. **Table IV** presents a comparison of message lengths for each step in the process. The proposed PQC-based KEMs are assessed using both ML-KEM and HQC algorithms. Steps involving encryption and decryption, including those that utilize encryption public keys, are marked with a green background for clarity.

The Falcon-512 algorithm produces significantly larger public keys and signatures than the ECDSA NIST P-256 algorithm. As a result, the message lengths in the proposed scheme are consistently longer than those specified in the IEEE standards. Nevertheless, because all certificate request messages are signed and verified using Falcon-512, the proposed method achieves NIST SL 1 and provides quantum-safe. Although the message lengths in the certificate request process exceed 1400 bytes, this does not affect V2I communication, which does not impose a 1400-byte size limit.

TABLE II.
THE CRYPTOGRAPHIC UNITS IN EACH MESSAGE BASED ON THE IEEE 1609.2 AND IEEE 1609.2.1 STANDARDS

| **IEEE 1609.2 and IEEE 1609.2.1** | |
|---|---|
| Message | Cryptographic Unit |
| EeEcaCertRequestSpdu | · ECC Public Key × 1<br>· ECDSA Signature × 1 |
| EcaEeCertResponseSpdu | · CA's Certificate × 4<br>· EC × 1<br>· ECDSA Signature × 1 |
| EeRaCertRequestSpdu | · ECC Public Key × 2<br>· AES Key × 2<br>· EC × 1<br>· ECDSA Signature × 1<br>· ECIES-based PK Recipient Info × 1 |
| RaEeCertAckSpdu | · RA's Certificate × 1<br>· ECDSA Signature × 1 |
| RaAcaCertRequestSpdu | · ECC Public Key × 2<br>· RA's Certificate × 1<br>· ECDSA Signature × 1 |
| AcaRaCertResponseSpdu | · Encrypted PC × 1<br>· ECIES-based PK Recipient Info × 1<br>· PCA's Certificate × 1<br>· ECDSA Signature × 2 |
| EeRaDownloadRequestSpdu | · EC × 1<br>· ECDSA Signature × 1<br>· ECIES-based PK Recipient Info × 1 |
| AcaEeCertResponsePrivateSpdu | · Encrypted PC × 1<br>· ECIES-based PK Recipient Info × 1<br>· PCA's Certificate × 1<br>· ECDSA Signature × 1 |
| Signed SPDU with BSM | · PC × 1<br>· ECDSA Signature × 1 |

TABLE III.
THE CRYPTOGRAPHIC UNITS IN EACH MESSAGE BASED ON THE PROPOSED GENERALIZED PSEUDONYM SCHEME

| **The Proposed Generalized Pseudonym Scheme** | |
|---|---|
| Message | Cryptographic Unit |
| EeEcaCertRequestSpdu | · PQC-based DSA Public Key × 1<br>· PQC-based DSA Signature × 1 |
| EcaEeCertResponseSpdu | · CA's Certificate × 4<br>· EC × 1<br>· PQC-based DSA Signature × 1 |
| EeRaCertRequestSpdu | · Encrypted DSA Public Key × 1<br>· EC × 1<br>· PQC-based DSA Signature × 1<br>· PQC-based KEM PK Recipient Info × 2 |
| RaEeCertAckSpdu | · RA's Certificate × 1<br>· PQC-based DSA Signature × 1 |
| RaAcaCertRequestSpdu | · RA's Certificate × 1<br>· PQC-based DSA Signature × 1<br>· PQC-based KEM PK Recipient Info × 1 |
| AcaRaCertResponseSpdu | · Encrypted PC × 1<br>· PQC-based KEM PreSharedKeyRecipientInfo × 1<br>· PQC-based DSA Signature × 2 |
| EeRaDownloadRequestSpdu | · EC × 1<br>· PQC-based DSA Signature × 1<br>· PQC-based KEM PK Recipient Info × 1 |
| AcaEeCertResponsePrivateSpdu | · Encrypted PC × 1<br>· PQC-based KEM PreSharedKeyRecipientInfo × 1<br>· PCA's Certificate × 1<br>· ECDSA Signature × 1 |
| Signed SPDU with BSM | · PC × 1<br>· DSA Signature × 1 |

TABLE IV.
THE COMPARISON OF MESSAGE LENGTHS (UNIT: BYTES)

| Message | IEEE Standards | | The Proposed Scheme | | | |
|---|---|---|---|---|---|---|
| | ECDSA P-256 + ECIES P-256 | | Falcon-512 + ML-KEM-512 | | Falcon-512 + HQC-128 | |
| | Length | SL | Length | SL | Length | SL |
| EeEcaCertRequestSpdu | 164 | 0 | 1629 | 1 | 1629 | 1 |
| EcaEeCertResponseSpdu | 954 | 0 | 8880 | 1 | 8880 | 1 |
| EeRaCertRequestSpdu | 416 | 0 | 3920 | 1 | 11378 | 1 |
| RaEeCertAckSpdu | 313 | 0 | 3146 | 1 | 4595 | 1 |
| RaAcaCertRequestSpdu | 628 | 0 | 4163 | 1 | 9341 | 1 |
| AcaRaCertResponseSpdu | 764 | 0 | 2444 | 1 | 2444 | 1 |
| EeRaDownloadRequestSpdu | 315 | 0 | 1553 | 1 | 5282 | 1 |
| AcaEeCertResponsePrivateSpdu | 504 | 0 | 3171 | 1 | 3171 | 1 |
| Signed SPDU including full PC with BSM | 212 | 0 | 813 | * | 813 | * |
| Signed SPDU including PC digest with BSM | 122 | 0 | 122 | * | 122 | * |

*: The SL of PC is 1, and the SL of signed SPDU with BSM is 0.

It is worth noting that although the message lengths of EeEcaCertRequestSpdu, EcaEeCertResponseSpdu, EeRaCertRequestSpdu, RaEeCertAckSpdu, RaAcaCertRequestSpdu, AcaRaCertResponseSpdu, EeRaDownloadRequestSpdu, and AcaEeCertResponsePrivateSpdu in the proposed scheme are longer than those defined in the current IEEE standards, these messages are all transmitted via V2I communication, where transmission resources are relatively less constrained.

In vehicular networks, V2V communication is more resource-limited, so the message length of the signed SPDUs with BSM requires closer attention. To achieve quantum-safe security, the proposed scheme employs a longer PCA signature within PCs. However, during vehicular communications, PC digests rather than full PCs are transmitted in most cases; a detailed explanation is provided in Appendix B.B. The PC digest adopts the HashedId8 format and requires only 8 bytes. Consequently, both the IEEE standards and the proposed scheme result in a Signed SPDU including PC digest with BSM of approximately 122 bytes. Within each transmission period, only one Signed SPDU including the full PC with BSM is transmitted, followed by $n_d$ transmissions of Signed SPDU including the PC digest with BSM. As a result, the proposed scheme yields message lengths for most Signed SPDUs with BSM that are consistent with those of the current IEEE standards, and transmission resources can be further conserved by adjusting the value of $n_d$.

Additionally, the length of the signed SPDU containing the BSM in the proposed scheme is approximately 813 bytes. This is within the 1400-byte limit, allowing the proposed approach to be applied to V2V communication while preserving quantum security for pseudonym certificates.

The HQC algorithm generates longer ciphertexts for the encapsulated key (i.e., the encapsulated KEK) compared to ML-KEM. Consequently, the message lengths for EeRaCertRequestSpdu and RaAcaCertRequestSpdu are longer when using the Falcon-512 and HQC-128 combination than when using Falcon-512 and ML-KEM-512. However, although AcaRaCertResponseSpdu and AcaEeCertResponsePrivateSpdu also include the ciphertext of the EE's PC, the proposed design adopts the PreSharedKeyRecipientInfo approach. This method uses the hash value of the pre-shared key, eliminating the need to transmit the encapsulated key itself. Therefore, the message lengths for these two types of messages remain the same regardless of whether ML-KEM-512 or HQC-128 is used.

According to SAE standard J2945/1_202004 [19] and the North American V2X Proof of Concept (PoC) implementation [23], BSMs are transmitted every 100 ms, while a Signed SPDU including a full PC with BSM is transmitted every 450 ms, corresponding to $n_d$ = 4. In practice, this results in one Signed SPDU including a full PC with BSM followed by four Signed SPDUs including a PC digest with BSM. Accordingly, each 0.5 second contains one Signed SPDU including a full PC with BSM and four SPDUs including a PC digest with BSM. Under these conditions, it can be observed that the total message size transmitted within 0.5 seconds under the IEEE standards is 700 bytes (1 × 212 bytes + $n_d$ × 122 byes), whereas the proposed scheme results in a total message size of 1301 bytes per second (1 × 813 bytes + $n_d$ × 122 byes). To address the larger packet sizes associated with PQC in future implementations, the difference can be reduced by increasing the value of $n_d$. In practice, this only requires adjusting a parameter, allowing EEs to implement the change efficiently.

### *D. The Comparison of Computation Time*

This section compares the cryptographic computation time required for each message involved in EC and PC requests between the IEEE 1609 series standards and the hybrid certificate-based pseudonym scheme proposed in this study. The comparison includes eight types of messages: EeEcaCertRequestSpdu, EcaEeCertResponseSpdu, EeRaCertRequestSpdu, RaEeCertAckSpdu, RaAcaCertRequestSpdu, AcaRaCertResponseSpdu, EeRaDownloadRequestSpdu, and AcaEeCertResponsePrivateSpdu, as detailed in **Table V** and **Table VI**. Additionally, the study analyzes the application of PCs in the transmission of ITS messages by generating a signed SPDU containing a BSM, and evaluates the computation time required at both the sender and receiver.

This study examines a typical EE responsible for transmitting BSMs, representing standard OBU functionality. It evaluates the complete process from the EE's EC and PC requests to the transmission of a signed SPDU containing the BSM. **Table VII** and **Table VIII** present a comparison of estimated computation times for each step in the process on Raspberry Pi 4 and Android. The proposed PQC-based KEMs are assessed using both ML-KEM and HQC algorithms. Steps involving key encapsulation and decapsulation, including those that utilize encryption public keys, are marked with a green background for clarity.

TABLE V.
THE CRYPTOGRAPHIC COMPUTATIONS IN EACH MESSAGE BASED ON THE IEEE 1609.2 AND IEEE 1609.2.1 STANDARDS

| **IEEE 1609.2 and IEEE 1609.2.1** | |
|---|---|
| Message | Cryptographic Computations |
| EeEcaCertRequestSpdu | · ECC Key Generation × 1<br>· ECDSA Signature Generation × 1 |
| EcaEeCertResponseSpdu | · ECDSA Signature Verification × 1<br>· ECDSA Signature Generation × 2 |
| EeRaCertRequestSpdu | · ECC Key Generation × 2<br>· ECDSA Signature Generation × 1<br>· Encryption Based on ECIES × 1 |
| RaEeCertAckSpdu | · Decryption Based on ECIES × 1<br>· ECDSA Signature Verification × 1<br>· ECDSA Signature Generation × 1 |
| RaAcaCertRequestSpdu | · KEF Computation × 2<br>· ECC Key Expansion × 2<br>· ECDSA Signature Generation × 1 |
| AcaRaCertResponseSpdu | · ECDSA Signature Verification × 1<br>· ECC Key Expansion × 1<br>· Encryption Based on ECIES × 1<br>· ECDSA Signature Generation × 3 |
| EeRaDownloadRequestSpdu | · ECDSA Signature Generation × 1<br>· Encryption Based on ECIES × 1 |
| AcaEeCertResponsePrivateSpdu | · ECDSA Signature Verification × 1<br>· Decryption Based on ECIES × 1 |
| Signed SPDU with BSM (sender) | · ECDSA Signature Generation × 1 |
| Signed SPDU with BSM (receiver) | · ECDSA Signature Verification × 2 |

TABLE VI.
THE CRYPTOGRAPHIC COMPUTATIONS IN EACH MESSAGE BASED ON THE PROPOSED GENERALIZED PSEUDONYM SCHEME

| **The Proposed Generalized Pseudonym Scheme** | |
|---|---|
| Message | Cryptographic Computations |
| EeEcaCertRequestSpdu | · PQC-based DSA Key Generation × 1<br>· PQC-based DSA Signature Generation × 1 |
| EcaEeCertResponseSpdu | · PQC-based DSA Signature Verification × 1<br>· PQC-based DSA Signature Generation × 2 |
| EeRaCertRequestSpdu | · DSA Key Generation × 1<br>· PQC-based DSA Signature Generation × 1<br>· Encryption Based on PQC-based KEM × 2 |
| RaEeCertAckSpdu | · Decryption Based on PQC-based KEM × 1<br>· PQC-based DSA Signature Verification × 1<br>· PQC-based DSA Signature Generation × 1 |
| RaAcaCertRequestSpdu | · PQC-based DSA Signature Generation × 1 |
| AcaRaCertResponseSpdu | · PQC-based DSA Signature Verification × 1<br>· Decryption Based on PQC-based KEM × 1<br>· Encryption Based on Pre-shared Key × 1<br>· PQC-based DSA Signature Generation × 3 |
| EeRaDownloadRequestSpdu | · PQC-based DSA Signature Generation × 1<br>· Encryption Based on PQC-based KEM × 1 |
| AcaEeCertResponsePrivateSpdu | · PQC-based DSA Signature Verification × 1<br>· Decryption Based on Pre-shared Key × 1 |
| Signed SPDU with BSM (sender) | · DSA Signature Generation × 1 |
| Signed SPDU with BSM (receiver) | · PQC-based DSA Signature Verification × 1<br>· DSA Signature Verification × 1 |

TABLE VII.
THE COMPARISON OF ESTIMATED COMPUTATION TIMES ON RASPBERRY PI 4 (UNIT: MILLISECONDS)

| Message | IEEE Standards | | The Proposed Scheme | | | |
|---|---|---|---|---|---|---|
| | ECDSA P-256 + ECIES P-256 | | Falcon-512 + ML-KEM-512 | | Falcon-512 + HQC-128 | |
| | Length | SL | Length | SL | Length | SL |
| EeEcaCertRequestSpdu | 3.948 | 0 | 31.275 | 1 | 31.275 | 1 |
| EcaEeCertResponseSpdu | 8.400 | 0 | 4.336 | 1 | 4.336 | 1 |
| EeRaCertRequestSpdu | 10.906 | 0 | 5.227 | 1 | 35.766 | 1 |
| RaEeCertAckSpdu | 7.457 | 0 | 2.753 | 1 | 25.867 | 1 |
| RaAcaCertRequestSpdu | 6.868 | 0 | 1.981 | 1 | 1.981 | 1 |
| AcaRaCertResponseSpdu | 17.143 | 0 | 7.257 | 1 | 30.370 | 1 |
| EeRaDownloadRequestSpdu | 6.581 | 0 | 2.523 | 1 | 17.792 | 1 |
| AcaEeCertResponsePrivateSpdu | 5.672 | 0 | 0.773 | 1 | 0.773 | 1 |
| Signed SPDU with BSM (sender) | 1.786 | 0 | 1.786 | * | 1.786 | * |
| Signed SPDU with BSM (receiver) | 9.657 | 0 | 5.203 | * | 5.203 | * |

*: The SL of PC is 1, and the SL of signed SPDU with BSM is 0.

TABLE VIII.
THE COMPARISON OF ESTIMATED COMPUTATION TIMES ON ANDROID (UNIT: MILLISECONDS)

| Message | IEEE Standards | | The Proposed Scheme | | | |
|---|---|---|---|---|---|---|
| | ECDSA P-256 + ECIES P-256 | | Falcon-512 + ML-KEM-512 | | Falcon-512 + HQC-128 | |
| | Length | SL | Length | SL | Length | SL |
| EeEcaCertRequestSpdu | 4.992 | 0 | 37.156 | 1 | 37.156 | 1 |
| EcaEeCertResponseSpdu | 11.781 | 0 | 6.185 | 1 | 6.185 | 1 |
| EeRaCertRequestSpdu | 19.100 | 0 | 6.718 | 1 | 28.671 | 1 |
| RaEeCertAckSpdu | 15.810 | 0 | 3.876 | 1 | 20.592 | 1 |
| RaAcaCertRequestSpdu | 8.205 | 0 | 2.936 | 1 | 2.936 | 1 |
| AcaRaCertResponseSpdu | 28.257 | 0 | 10.327 | 1 | 27.043 | 1 |
| EeRaDownloadRequestSpdu | 13.852 | 0 | 3.515 | 1 | 14.491 | 1 |
| AcaEeCertResponsePrivateSpdu | 13.442 | 0 | 0.940 | 1 | 0.940 | 1 |
| Signed SPDU with BSM (sender) | 2.368 | 0 | 2.368 | * | 2.368 | * |
| Signed SPDU with BSM (receiver) | 14.089 | 0 | 7.358 | * | 7.358 | * |

*: The SL of PC is 1, and the SL of signed SPDU with BSM is 0.

The key generation time of Falcon-512 is significantly longer than that of ECDSA NIST P-256, resulting in increased computational time for generating the EeEcaCertRequestSpdu message. However, the signature generation time of Falcon-512 is comparable to that of ECDSA NIST P-256, and its signature verification is faster. Therefore, Falcon-512 demonstrates greater efficiency in the signing and verification processes. In addition, the key encapsulation and decapsulation times of ML-KEM-512 are shorter than those of ECIES NIST P-256, leading to reduced computational time when encryption and decryption are performed using ML-KEM-512. Conversely, HQC exhibits longer computation times for key encapsulation and decapsulation, resulting in increased processing time when encryption and decryption are performed using HQC.

Another notable aspect of the proposed approach is the optimization of the RaAcaCertRequestSpdu message. Since it does not require cocoon public key expansion, the processing time is significantly reduced. Furthermore, for the KEM-based encrypted SPDU for responses, the proposed method utilizes the PreSharedKeyRecipientInfo structure. This eliminates the need to recalculate key encapsulation and decapsulation during the processing of the AcaEeCertResponsePrivateSpdu message, thereby improving processing efficiency. The proposed method achieves quantum-safe in both the certificate request process and the issued certificates, offering stronger security guarantees compared to those defined in IEEE 1609.2 and IEEE 1609.2.1.

Regarding the use of pseudonym certificates in transmitting the SPDUs with BSMs, the computational time required on the sender remains consistent, as all messages are signed using ECDSA NIST P-256. However, on the receiver, although the BSM is signed with ECDSA NIST P-256, the pseudonym certificate embedded within the SPDU is verified using Falcon-512. Since Falcon-512 provides faster verification performance than ECDSA NIST P-256, the proposed method demonstrates improved efficiency on the receiver.

To evaluate the feasibility of the proposed approach under varying traffic congestion conditions, this study adopts the six Levels of Service (LOS) defined in the *Highway Capacity Manual, 7th Edition: A Guide for Multimodal Mobility Analysis* published by the National Academies Press [34]. The corresponding vehicle densities are converted into units of vehicles per kilometer per lane. Assuming a V2X communication range of 300 meters [35] and that each EE transmits 10 SPDUs per second [19], the number of SPDUs per second under different LOS conditions can be derived, as summarized in **Table IX**. For example, Level A corresponds to a density of 11 vehicles/mile/lane (approximately 6.835 vehicles/km/lane). Under a 300 m communication range and a transmission rate of 10 SPDUs per vehicle per second, this results in an average of 20.505 SPDUs. For Level F, the density exceeds 45 vehicles/mile/lane (approximately 27.962 vehicles/km/lane). Since such density cannot increase indefinitely in physical terms, this study assumes an effective vehicle length of 30 meters (i.e., vehicle length including safe spacing) under congested conditions [34], corresponding to approximately 33.333 vehicles per kilometer. Under these conditions, with a 300 m communication range and a transmission rate of 10 SPDUs per EE per second, the average number of SPDUs is approximately 100.

TABLE IX.
THE NUMBER OF SPDUS PER EE PER SECOND UNDER DIFFERENT OF LOS

| LOS | vehicle/mile/lane | vehicle/km/lane | The number of SPDUs per EE per second‡ |
|---|---|---|---|
| A | 11 | 6.835 | 20.505 |
| B | 18 | 11.185 | 33.554 |
| C | 25 | 15.534 | 46.603 |
| D | 35 | 21.748 | 65.244 |
| E | 45 | 27.962 | 83.885 |
| F | > 45 | 33.333† | 100.000 |

†: Assumes 33.333 vehicles per kilometer, calculated as 1000 m divided by an effective vehicle length of 30 m [34].
‡: Assumes a V2X communication range of 300 m [35], with each EE transmitting 10 SPDUs per second.

The analysis focuses on whether the requirements can be met under Level F conditions, with the understanding that if Level F is satisfied, all other LOS conditions can also be met. Under these conditions, each sender is required to sign 10 Signed SPDUs with BSM per second, while each receiver must verify 100 Signed SPDUs with BSM per second. Using the computation times reported in **Tables VII** and **VIII**, the number of SPDUs that can be processed per second is calculated to evaluate computational performance, as summarized in **Tables X** and **XI**. For example, on a Raspberry Pi 4, signing a single Signed SPDU with BSM takes 1.786 ms, allowing up to 559.910 Signed SPDUs with BSM to be signed per second, which exceeds the requirement of 10 per second. In contrast, verifying a Signed SPDU with BSM on an Android-based EE under the IEEE standards requires 14.089 ms, permitting only 70.977 verifications per second, which is below the required 100, indicating that IEEE standards would be unable to handle high-density traffic conditions. By comparison, the proposed scheme meets the computational requirements on both Raspberry Pi 4 and Android across all LOS conditions. Therefore, the proposed scheme is better suited for real-world V2X environments and can support the computational demands of high-density traffic scenarios.

TABLE X.
COMPARISON OF COMPUTATIONAL PERFORMANCE ON RASPBERRY PI 4 UNDER LEVEL F CONDITIONS

| Message | Requirement | IEEE Standards | | The Proposed Scheme | |
|---|---|---|---|---|---|
| | | The number of SPDUs | Check | The number of SPDUs | Check |
| Signed SPDU with BSM (sender) | 10 | 559.910 | Pass | 559.910 | Pass |
| Signed SPDU with BSM (receiver) | 100 | 103.552 | Pass | 192.197 | Pass |

TABLE XI.
COMPARISON OF COMPUTATIONAL PERFORMANCE ON ANDROID UNDER LEVEL F CONDITIONS

| Message | Requirement | IEEE Standards | | The Proposed Scheme | |
|---|---|---|---|---|---|
| | | The number of SPDUs | Check | The number of SPDUs | Check |
| Signed SPDU with BSM (sender) | 10 | 422.297 | Pass | 422.297 | Pass |
| Signed SPDU with BSM (receiver) | 100 | 70.977 | **Fail** | 135.906 | Pass |

*E. Security Discussions*

This section discusses the proposed scheme from two perspectives: the scenario of attacks on the infrastructure and the anonymity of EEs.

**1). The Scenario of Attacks on the Infrastructure**

Certificates used by current V2X infrastructure devices generally have long validity periods, as shown in **Table XII**. If an attacker compromises infrastructure-related devices, the potential impact is significant. For example, compromising the RCA would allow the attacker to forge RCA signatures and use RCA certificates, effectively gaining control over the entire V2X SCMS.

TABLE XII.
CERTIFICATE VALIDITY PERIODS AND VULNERABILITY TO QUANTUM COMPUTING ATTACKS

| Certificate | Validity Period | IEEE Standards (Resistant to Quantum Attacks) | The Proposed Scheme (Resistant to Quantum Attacks) |
|---|---|---|---|
| RCA | 70 years [5] | **Fail** | Pass |
| ICA | 15 years [7] | **Fail** | Pass |
| ECA | 10 years [7] | **Fail** | Pass |
| PCA | 6 years [7] | **Fail** | Pass |
| RA | 6 years [7] | **Fail** | Pass |
| EC | 6 years [7] | **Fail** | Pass |
| PC | 1 week [7] | Pass | Pass |

In recent years, some studies have explored the use of Shor's algorithm to break ECC [36]; however, no reliable or rigorous experimental results have yet been reported. As a result, it remains difficult to precisely estimate the computational time required to break ECC. This study therefore refers to authoritative literature from *Physical Review Letters* and assumes that breaking ECC using NIST P-256 would require the same time as breaking RSA-2048, approximately 177 days [37]. Under this assumption, if ECC is used and the certificate validity exceeds 177 days, the certificates could be vulnerable to quantum attacks, potentially allowing an attacker to recover the private key and forge signatures. Consequently, using IEEE standards, RCA, ICA, ECA, PCA, and RA certificates within the infrastructure environment may be insecure, and even ECs for EEs could be compromised. However, since PC for EEs have a validity of only one week, and quantum computation would require 177 days to extract the private key, any attack on a PC would be ineffective because the certificate would have already expired, as shown in **Table XII**. This indicates that short-lived certificates for EEs can mitigate the risk of quantum attacks. For infrastructure certificates such as RCA, ICA, ECA, PCA, and RA, the proposed scheme employs PQC. Currently, no literature provides evidence that quantum computers can break ML-KEM or Falcon, ensuring that these certificates are quantum-secure.

The proposed scheme is designed to provide quantum security by employing pure PQC certificates with long validity requirements in the infrastructure environment, including RCA's certificates, ICA's certificates, ECA's certificates, PCA's certificates, RA's certificates, and EE's ECs. In contrast, EE's PCs have short validity periods. Even with potential future improvements in quantum computing efficiency, security can be maintained by further reducing the validity period of PCs, as illustrated in **Table XII**.

**2). The Anonymity of EEs**

The proposed scheme primarily ensures quantum-safe security for V2I communications. This section analyzes a scenario that assumes a man-in-the-middle (MITM) attack in which the adversary is equipped with quantum computing capabilities. It is assumed that current quantum computers are still unable to solve the Shortest Vector Problem (SVP) and the Learning With Errors (LWE) problem in polynomial time; that is, lattice-based cryptography (e.g., ML-KEM, ML-DSA, and Falcon) remains secure against quantum attacks.

Under the assumption of a MITM attack between the EE and the ECA, the ECA is pre-provisioned with the Falcon-based canonical public key of the EE. The EeEcaCertRequestSpdu sent by the EE carries a signature generated using the Falcon-based canonical private key. Consequently, the ECA can verify this signature using the corresponding Falcon-based canonical public key and confirm that the message was indeed sent by a trusted EE. Although an attacker may intercept the EeEcaCertRequestSpdu, any modification to its contents would invalidate the signature, and the attacker would be unable to forge a valid Falcon-based signature. Similarly, the EcaEeCertResponseSpdu includes a Falcon-based signature generated by the ECA, and an attacker cannot successfully alter the message without being detected. This mechanism ensures the security of communication between the EE and the ECA.

Under the assumption of a MITM attack between the EE and the RA, message exchanges are protected not only by Falcon signatures but also by the encryption based on ML-KEM encryption. As a result, the attacker cannot break ML-KEM and therefore cannot obtain the plaintext of the transmitted messages.

Under the assumption of a MITM attack between the RA and the PCA, all messages exchanged between them, including RaAcaCertRequestSpdu and AcaRaCertResponseSpdu, carry Falcon-based signatures generated by the respective parties. Consequently, although an attacker may intercept these messages, any modification to the contents of RaAcaCertRequestSpdu or AcaRaCertResponseSpdu would invalidate the corresponding Falcon-based signatures, which the attacker is unable to forge. This mechanism ensures the security of communication between the RA and the PCA.

In this study, privacy protection is defined as the inability to derive a linkage between the ECDSA public key in the EE's PC and the Falcon-based public key in the EE's EC. Prior to transmission, the EE encrypts its ECDSA public key using the PCA's ML-KEM public key and packages it within the EeRaCertRequestSpdu, which is then sent to the RA. Because the RA cannot decrypt the ciphertexts based on ML-KEM, it cannot obtain the plaintexts of the EE's ECDSA public key contained in the EeRaCertRequestSpdu. Furthermore, the EE's PC issued by the PCA is packaged in the AcaEeCertResponsePrivateSpdu, which is encrypted using a pre-shared key. As a result, the RA cannot access the plaintexts of PC or the ECDSA public key contained within it.

Furthermore, even in the presence of a MITM attack, the adversary cannot break the encryption based on ML-KEM and therefore cannot derive any association between the ECDSA public key in the EE's PC and the Falcon-based public key in the EE's EC. This design effectively preserves the privacy of the EE.

This study also compares the entropy values of encrypted ECDSA public keys under different security strengths. Specifically, at the 128-bit security level, an ECDSA NIST P-256 public key is encrypted using ML-KEM-512; at the 192-bit security level, an ECDSA NIST P-384 public key is encrypted using ML-KEM-768; and at the 256-bit security level, an ECDSA NIST P-521 public key is encrypted using ML-KEM-1024. For each configuration, 1,000 executions are performed, and the minimum entropy value (i.e., the worst case) is selected for evaluation and comparison. The experimental results are presented in **Table XIII**. A binomial test is applied to the entropy values, yielding p-values that are all greater than the threshold of 0.000005. Consequently, the results pass the randomness validation specified in NIST SP 800-90B [38].

TABLE XIII.
ENTROPY VALIDATION P-VALUES

| Security Strength | KEM | ECDSA | Entropy on Raspberry Pi 4 | Entropy on Android |
|---|---|---|---|---|
| 128 | ML-KEM-512 | NIST P-256 | 0.00496 | 0.00043 |
| 192 | ML-KEM-768 | NIST P-384 | 0.00011 | 0.00057 |
| 256 | ML-KEM-1024 | NIST P-521 | 0.00291 | 0.00352 |

### *F. Migration Issues and Costs*

This section discusses the migration costs from the perspective of the roles in the SCMS. The proposed scheme primarily enhances the quantum-safe security of V2I communications and the supporting infrastructure. As a result, only the RCA, ICA, ECA, PCA, and RA need to support pure PQC, meaning that, from an implementation cost perspective, modifications are required for only these five devices. PKI providers may choose to scale the number of infrastructure devices based on market size and redundancy requirements, but the minimum required number remains the five devices mentioned above.

The current IEEE standard already supports crypto agility through the BasePublicEncryptionKey and PublicVerificationKey structures. In this study, PQC is integrated by adding the structure of MLKEM512PublicKey to the CHOICE within the structure of BasePublicEncryptionKey (as shown in **Fig. 12**) and the structure of Falcon512PublicKey to the CHOICE within the structure of PublicVerificationKey (as shown in **Fig. 13**). This approach enables support for post-quantum cryptography while maintaining compatibility with existing systems.

```
BasePublicEncryptionKey ::= CHOICE {
  eciesNistP256           EccP256CurvePoint,
  eciesBrainpoolP256r1    EccP256CurvePoint,
  …,
  ecencSm2                EccP256CurvePoint,
  mlkem512                MLKEM512PublicKey
}
```

**Fig. 12.** The structure of BasePublicEncryptionKey.

```
PublicVerificationKey ::= CHOICE {
  ecdsaNistP256           EccP256CurvePoint,
  ecdsaBraIoolP256r1      EccP256CurvePoint,
  … ,
  ecdsaBrainpoolP384r1    EccP384CurvePoint,
  ecdsaNistP384   EccP384CurvePoint,
  ecsigSm2                EccP256CurvePoint,
  falcon512               Falcon512PublicKey
}
```

**Fig. 13.**

**Fig. 13.** The structure of PublicVerificationKey.

Subsequent newly manufactured EEs can adopt the proposed structures of BasePublicEncryptionKey and PublicVerificationKey in this study to support PQC, while still remaining capable of interpreting the original elliptic-curve-based BasePublicEncryptionKey and PublicVerificationKey structures. Furthermore, since these functions primarily involve public-key applications (e.g., encryption and verification), they can be implemented in software and updated via Over-the-Air (OTA) technology, eliminating the need to physically recall and modify each EE at the factory.

In summary, the proposed scheme requires only minimal modifications to infrastructure devices, enables EEs to be updated through OTA, maintains compatibility with existing systems, and therefore entails relatively low migration costs.

## V. Conclusions and Future Work

Given the challenges faced by vehicular communications, including packet length limitation, signature generation and verification efficiency, SL, and protection of vehicular privacy, the pseudonym scheme proposed in this study based on hybrid certificates offers a promising transitional solution toward PQC. This approach enables infrastructure devices to remain resilient against quantum attacks while simultaneously addressing the key limitations of current vehicular communications.

- **Packet Length Limitation**: The proposed hybrid certificate design combines PQC and ECC. Specifically, the CA employs the Falcon-512 algorithm, while EEs use the ECDSA P-256 algorithm. Under this configuration, the total protocol data unit size remains below the 1400-byte threshold, which satisfies the constraints of V2V communication.
- **Signature Generation and Verification Efficiency**: The implementation of various PQC signature algorithms was evaluated on a resource-constrained EE. Experimental results indicate that Falcon-512 achieves shorter verification times compared to the widely used ECDSA P-256, demonstrating its practical efficiency in real-world deployment.
- **SL**: Current industry standards rely on ECC, which is considered vulnerable in the face of quantum computing. The proposed hybrid certificate scheme meets the security levels defined by the NIST. In this framework, Falcon-512 is used as the DSA, while ML-KEM-512 is employed for KEM, thereby satisfying quantum-safe requirements.
- **Vehicle Privacy**: To preserve user privacy, the proposed pseudonym scheme ensures that the PCA, RA, and other EEs are unable to infer any link between a enrollment public key and the public key contained in a PC. This unlinkability ensures the confidentiality of the EE's identity. Additionally, EEs are allowed to request multiple PCs and rotate their usage over time to further strengthen privacy protection.

The pure PQC certificate proposed in this study provides comprehensive protection for the infrastructure devices in the SCMS, including certificates issued by the CAs, the RAs, and the ECs of EEs. In contrast, the hybrid certificate scheme, which combines PQC with ECC, secures the PCs used by EEs. However, the SPDUs in the hybrid approach continue to rely on ECC-based signatures. To mitigate the associated risks, frequent updates of PCs are recommended to reduce the likelihood of successful attacks.

Furthermore, NIST is currently evaluating additional digital signature algorithms that offer shorter signature lengths and faster performance. Once such algorithms are standardized, future deployments can adopt these improved algorithms to construct pure PQC-based PCs for EEs. This would ensure that all components of the SCMS are protected under PQC, thereby offering comprehensive resilience against quantum-enabled adversaries.

## Acknowledgments

The RCA developed in this study has successfully passed the interoperability testing conducted at the OmniAir plugfest conference. It has also been included in the Test CTL published by the SCMS Manager in the United States, which is available at: https://www.scmsmanager.org/publications/. In addition, the RCA certificate has been incorporated into the European Certificate Trust List (ECTL), maintained by the C-ITS Point of Contact (CPOC) of the European Union, accessible at: https://cpoc.jrc.ec.europa.eu/ECTL.html. This study further acknowledges the support of the NIST for providing the Automated Cryptographic Validation Testing System (ACTVS). This platform enabled the automated validation of cryptographic algorithms, including PQC algorithms such as ML-KEM, ML-DSA, and SLH-DSA, as well as the AES and SHA. All algorithms tested through ACTVS received passing validation results. The authors would like to express their sincere appreciation to the IEEE 1609 Working Group for adopting the proposed contributions and incorporating the relevant content into IEEE 1609.2-2025/D5 and IEEE 1609.2.1-2025/D3. The authors also extend their gratitude to ETSI ITS Working Group 5 for accepting the proposals and including the corresponding content in the forthcoming versions of ETSI TS 103 097 and ETSI TS 102 941. The implementation of this study was partially supported through the contributions of colleagues J.-F. Lin and P.-Y. Su. The contributions and support provided by OmniAir, the SCMS Manager, the CPOC, NIST, and my colleagues are sincerely appreciated.

## APPENDIX A

The used acronyms in this study are summarized as follows.

| | |
|---|---|
| **AC** | Authorization Certificate |
| **ACVTS** | Automated Cryptographic Validation Testing System |
| **AES** | Advanced Encryption Standard |
| **BKE** | Butterfly Key Expansion |
| **BSM** | Basic Safety Message |
| **C-ITS** | Cooperative Intelligent Transport Systems |
| **CA** | Certificate Authority |
| **CCF** | Certificate Chain File |
| **CCMS** | C-ITS Security Credential Management System |
| **CPOC** | C-ITS Point of Contact |
| **CRYSTALS** | Cryptographic Suite for Algebraic Lattices |
| **CTL** | Certificate Trust List |
| **DEK** | Data-Encryption-Key |
| **DSA** | Digital Signature Algorithm |
| **EC** | Enrollment Certificate |
| **ECA** | Enrollment Certificate Authority |
| **ECC** | Elliptic-Curve Cryptography |
| **ECDH** | Elliptic-Curve Diffie–Hellman |
| **ECDLP** | Elliptic-Curve Discrete Logarithm Problem |
| **ECDSA** | Elliptic-Curve Digital Signature Algorithm |
| **ECIES** | Elliptic-Curve Integrated Encryption Scheme |
| **ECTL** | European Certificate Trust List |
| **EE** | End Entity |
| **ETSI** | European Telecommunications Standards Institute |
| **HMAC** | keyed-Hash Message Authentication Code |
| **HQC** | Hamming Quasi-Cyclic |
| **I2I** | Infrastructure-to-Infrastructure |
| **ICA** | Intermediate Certificate Authority |
| **ITS** | Intelligent Transportation System |
| **KDF** | Key Derivation Function |
| **KEF** | Key Expansion Function |
| **KEK** | Key-Encryption-Key |
| **KEM** | Key Encapsulation Mechanism |
| **LOS** | Levels of Service |
| **LWE** | Learning With Errors |
| **MITM** | Man-In-The-Middle |
| **ML-DSA** | Module-Lattice-Based DSA |
| **ML-KEM** | Module-Lattice-Based KEM |
| **NIST** | National Institute of Standards and Technology |
| **OBU** | On-Board Unit |
| **OTA** | Over-the-Air |
| **PC** | Pseudonym Certificate |
| **PCA** | Pseudonym Certificate Authority |
| **POC** | Proof of Concept |
| **PS** | Pseudonym Scheme |
| **PSID** | Provider Service IDentifier |
| **RA** | Registration Authority |
| **RCA** | Root Certificate Authority |
| **PKI** | Public Key Infrastructure |
| **RSU** | Road-Side Unit |
| **SCMS** | Security Credential Management System |
| **SHA** | Secure Hash Algorithm |
| **SL** | Security Level |
| **SLH-DSA** | Stateless Hash-Based DSA |
| **SPaT** | Signal Phase and Timing |
| **SPDU** | Secure Protocol Data Unit |
| **SVP** | Shortest Vector Problem |
| **TBSC** | To-Be-Signed-Certificate |
| **TBSD** | To-Be-Signed-Data |
| **TBSR** | To-Be-Signed-Request |
| **V2I** | Vehicle-to-Infrastructure |
| **V2V** | Vehicle-to-Vehicle |
| **V2X** | Vehicle-to-Everything |
| **VKI** | Verify Key Indicator |
| **WAVE** | Wireless Access in the Vehicular Environment |
| **WSM** | WAVE Short Message |

## APPENDIX B

To provide implementation details, Appendix B presents literature related to the authors' work for readers interested in further study. It is noted that these are not references and do not constitute self-citations.

### *A. Implementation and Interoperability Testing*

The SCMS implemented in this study has participated in multiple OmniAir Plugfest events from 2023 to the present. Interoperability testing has been conducted with OBUs and RSUs from multiple vendors, including UNEX and Clientron, following both IEEE and ETSI standards. Detailed implementation information and OBU specifications can be found in [b1] and [23]. The some certificates developed in this study have also been included in the US SCMS Manager's Test Certificate Trust List and the European Certificate Trust List (ECTL) maintained by the C-ITS Point of Contact (CPOC), as acknowledged in this work.

The preliminary concept of the hybrid certificate proposed in this study was presented at the 2023 OmniAir Plugfest under the title "Challenges in the Practical Application of Post-Quantum Cryptography in Security Credential Management System" [b2]. OmniAir Plugfest is a key interoperability testing event in the vehicular networking domain, attended annually by numerous equipment vendors and PKI providers. Furthermore, partial implementation details of the hybrid certificate were published in 2024 at an IEEE conference [23].

This study constitutes an extension of the conference paper [23]. In the prior work [23], the authors discussed only the hybrid certificate format, whereas the certificate issuance process and the integration of ML-KEM and HQC for encryption are original contributions of this study.

### *B. Practical Deployment and Operation*

Current vehicular network deployment and operations are primarily based on parameters from the US POC. Most V2V SPDUs transmit only PC digests rather than full PCs to conserve V2V communication resources; operational details are provided in [b3]. Furthermore, [b3] discusses the probability of HashedId8 collisions, with detailed proofs provided therein.

### C. Butterfly Key Expansion and Certificate Privacy

In the IEEE standards, BKE relies on ECC and is based on the ECDLP. As long as ECDLP cannot be solved in polynomial time, deriving the original public key from the expanded key would require non-polynomial time for classic computers, thereby protecting privacy; detailed proofs are provided in [25]. In contrast, this study encrypts the public key intended for the PC directly using the PCA's ML-KEM public key, providing privacy protection based on the SVP and LWE problem, thus achieving quantum-safe security. Furthermore, the PCs are short-lived, with an implementation validity period set to seven days. Within a single period, the PCA issues multiple PCs to the same EE, with the EE switching to a new PC every five minutes, further enhancing privacy [b3],[25].

### D. Related Publication List

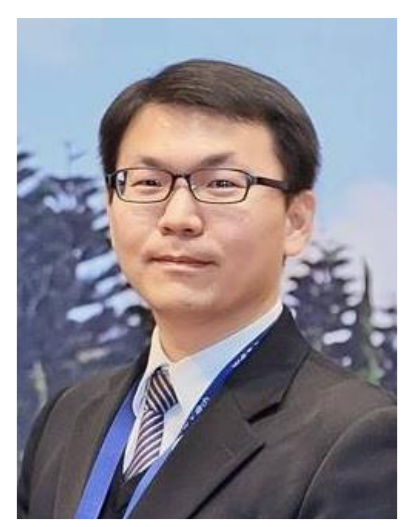

**Abel C. H. Chen** (Senior Member, IEEE) has published over 400 journal articles, conference papers, standards, and patents. His contributions were published in *IEEE Transactions on Intelligent Transportation Systems*, *IEEE Internet of Things Journal, ACM Transactions on Sensor Networks*, *Information Science*, and so on. He was listed among the Top 2% Scientists Worldwide from 2022 to 2025 by Stanford University. Some of his publications have been recognized as highly cited papers on Web of Science using data from Essential Science Indicators (ESI). He serves as the editor for several journals such as *Scientific Data*, *IEEE Open Journal of Intelligent Transportation Systems*, and *Network: Computation in Neural Systems*. He also served as an associate editor for several journals (e.g., *IEEE Access*, *IEICE Transactions on Information and Systems*, and so on.) and the chair for several conferences (e.g., *WWW'21 Workshop*, *IEEE TrustCom 2021 Workshop*, *IEEE APNOMS 2020*, and so on *IEEE ICC 2020*).

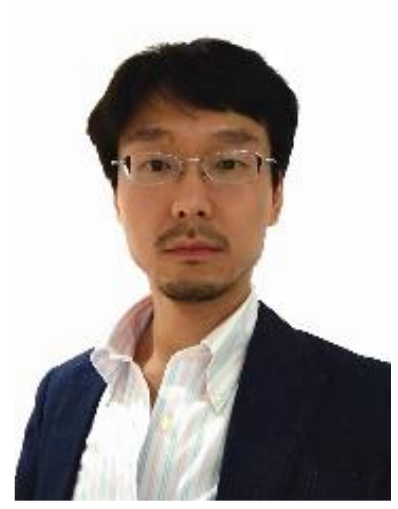

**F. J. Hwang** is a professor, the leading PI of the Intelligent Logistics and Transport Group, at the Department of Business Management, National Sun Yat-sen University, Taiwan. His research interests center around data-driven optimization, intelligent transportation and logistics, management information science, and computational intelligence. F.J. has published regularly in the leading journals, including *Annals of Operations Research*, *Computational Optimization and Applications*, *Journal of the Operational Research Society*, *Journal of Scheduling*, *Computers & Operations Research*, *Discrete Optimization*, *Engineering Optimization*, *Computers & Industrial Engineering*, *IEEE Transactions on Intelligent Transportation System*, *International Journal of Sustainable Transportation*, *ACM Transactions on Sensor Networks*, etc. He has been serving as the Guest Editor for more than ten Q1/Q2 SCI journals, including *Journal of Global Information Management*, *Journal of Traffic and Transportation Engineering*, *Journal of Database Management*, *Mobile Information Systems*, etc., as well as the General/Session Chair for more than ten international conferences/workshops and invited to give more than 40 research talks and conference presentations.

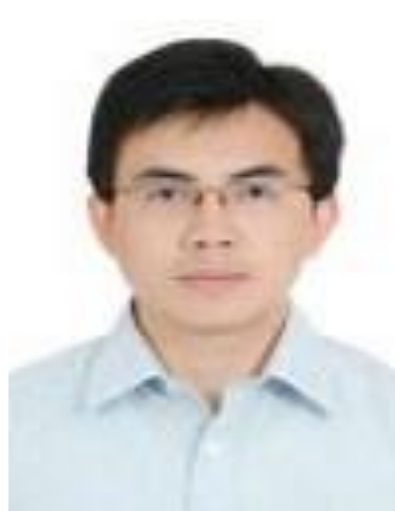

**Yu-Chih Wei** is an associate professor in the Department of Information and Finance Management at the National Taipei University of Technology. He holds a Ph.D. in Information Management from National Central University. His research interests include FinTech security, ISRA, RegTech, vehicular ad-hoc network (VANET) security, information security management, and business continuity management. Before pursuing an academic career, Dr. Wei was a researcher at the Information & Communication Security Laboratory of Chunghwa Telecom Co., Ltd.

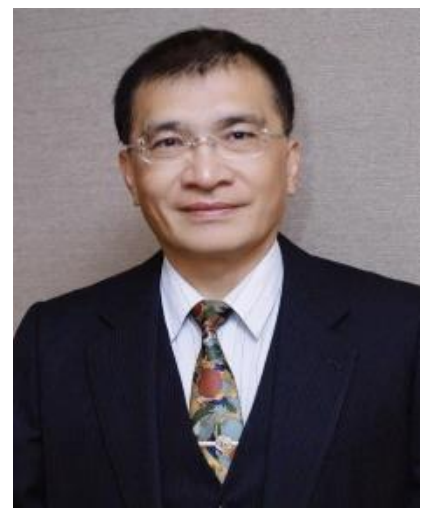

**Chin-Chen Chang** (Fellow, IEEE) has worked on many different topics in information security, cryptography, multimedia image processing and published several hundreds of papers in international conferences and journals and over 40 books. He was cited over 52,690 times and has an h-factor of 105 according to Google Scholar. Several well-known concepts and algorithms were adopted in textbooks. He also worked with the National Science Council, Ministry of Technology, Ministry of Education, Ministry of Transportation, Ministry of Economic Affairs and other Government agencies on more than 100 projects and holds 40 patents, including one in US and two in China. He served as Honorary Professor, Consulting Professor, Distinguished Professor, and Guest Professor at over 72 academic institutions and received Distinguished Alumni Award's from his Alma Master's. He also served as Editor or Chair of several international journals and conferences and had given almost a thousand invited talks at institutions including Chinese Academy of Sciences, Academia Sinica, Tokyo University, Kyoto University, National University of Singapore, Nanyang Technological University, The University of Hong Kong, National Taiwan University and Peking University. Professor Chang has mentored 7 postdoctoral, 72 PhD students and 209 master students, most of whom hold academic positions at major national or international universities. He has been the Editor-in-Chief of Information Education, a magazine that aims at providing educational materials for middle-school teachers in computer science. He is a leader in the field of information security of Taiwan. He founded the Chinese Cryptography and Information Security Association, accelerating information security the application and development and consulting on the government policy. He is also the recipient of several awards, including the Top Citation Award from Pattern Recognition Letters, Outstanding Scholar Award from Journal of Systems and Software, and Ten Outstanding Young Men Award of Taiwan. He was elected as a Fellow of IEEE in 1998, a Fellow of IET in 2000, a Fellow of CS in 2020, an AAIA Fellow in 2021, a Member of the Academy of Europe (AE) in 2022, and a Member of the European Academy of Sciences and Arts (EASA) in the same year for his contribution in the area of information security.

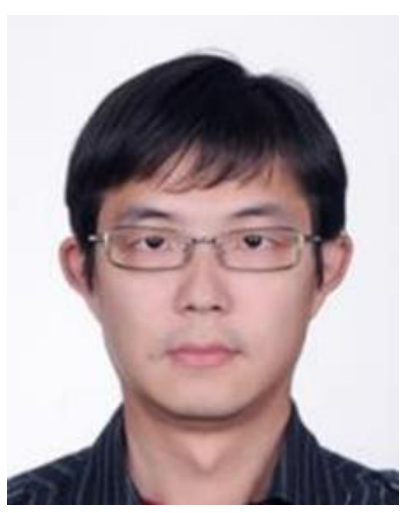

**Bon-Yeh Lin** received his Ph.D. degree in information management from National Chiao Tung University in 2011. He is a senior research fellow for Chunghwa Telecom Laboratories. His contributions were published in Enterprise Information Systems, International Journal of Mobile Communications, International Journal of Network Management, and so on. His research interests include network security and public key infrastructure.